\documentclass[useAMS,usenatbib]{mn2e}
\usepackage{epsfig}
\usepackage{epstopdf}
\usepackage{amsmath}
\usepackage{amssymb}
\usepackage{natbib}
\newcommand{\deriv}{{\rm d}}

\newcommand{\HI}{{\rm H\textsc{i}}}
\newcommand{\dd}{\ensuremath{{\rm d}}}

\newcommand{\cm}{\ensuremath{{\rm cm}}}

\newcommand{\mpc}{\ensuremath{{\rm Mpc}}}
\newcommand{\erg}{\ensuremath{{\rm erg}}}
\newcommand{\Hz}{\ensuremath{{\rm Hz}}}
\newcommand{\km}{\ensuremath{{\rm km}}}
\newcommand{\s}{\ensuremath{{\rm s}}}
\newcommand{\kms}{\ensuremath{\km\,\s^{-1}}}
\newcommand{\yr}{\ensuremath{{\rm yr}}}

\newcommand{\K}{\ensuremath{{\rm K}}}
\newcommand{\msun}{\ensuremath{{M_{\odot}}}}

\newcommand{\Gloc}{\ensuremath{\Gamma_{\rm local}}}

\newcommand{\Ls}{\ensuremath{L_{\rm s}}}

\newcommand{\NHI}{\ensuremath{N_{\HI}}}
\newcommand{\nHI}{\ensuremath{n_{\HI}}}
\newcommand{\nH}{\ensuremath{n_{\rm H}}}
\newcommand{\xHI}{\ensuremath{x_{\HI}}}
\newcommand{\NLLS}{\ensuremath{N_{\rm LLS}}}
\newcommand{\Di}{\ensuremath{\Delta_{\rm i}}}
\newcommand{\Dvir}{\ensuremath{\Delta_{\rm vir}}}

\newcommand{\rLLS}{\ensuremath{r_{\rm LLS}}}
\newcommand{\xLLS}{\ensuremath{x_{\rm LLS}}}
\newcommand{\rvir}{\ensuremath{r_{\rm vir}}}
\newcommand{\rsp}{\ensuremath{r_{\rm sp}}}
\newcommand{\rs}{\ensuremath{r_{\rm s}}}
\newcommand{\cvir}{\ensuremath{c_{\rm vir}}}
\newcommand{\ra}{\ensuremath{r_{\rm a}}}
\newcommand{\sa}{\ensuremath{\sigma_{\rm a}}}
\newcommand{\na}{\ensuremath{n_{\rm a}}}
\newcommand{\ns}{\ensuremath{n_{\rm s}}}
\newcommand{\nss}{\ensuremath{n_{\rm ss}}}
\newcommand{\mfp}{\ensuremath{{\lambda_{\rm mfp}}}}
\newcommand{\mabs}{\ensuremath{{M_{\rm min}^{\rm abs}}}}
\newcommand{\sfr}{\ensuremath{{\dot{M}_{\rm \star}}}}
\newcommand{\macc}{\ensuremath{M_{\rm acc}}}
\newcommand{\memit}{\ensuremath{{M_{\rm min}^{\rm emit}}}}
\newcommand{\fesc}{\ensuremath{{f_{\rm esc}}}}
\newcommand{\lion}{\ensuremath{{l_{\rm ion}}}}
\newcommand{\eion}{\ensuremath{{\epsilon_{\rm 912}}}}
\newcommand{\mhalo}{\ensuremath{M_{\rm halo}}}
\newcommand{\mf}{\ensuremath{M_{\rm f}}}
\newcommand{\zrei}{\ensuremath{z_{\rm rei}}}
\newcommand{\Yhe}{\ensuremath{Y_{\rm He}}}
\newcommand{\fagn}{\ensuremath{f_{\rm AGN}}}
\def\gsim{\;\rlap{\lower 2.5pt
 \hbox{$\sim$}}\raise 1.5pt\hbox{$>$}\;}
\def\lsim{\;\rlap{\lower 2.5pt
   \hbox{$\sim$}}\raise 1.5pt\hbox{$<$}\;}

\begin{document}

\title[Evolution of the Ionizing Background]{The Flatness and Sudden Evolution of the Intergalactic Ionizing Background}

\author[J.~A.~Mu{\~n}oz {\it et al.}]{
Joseph~A.~Mu{\~n}oz$^1$\thanks{jamunoz@physics.ucsb.edu}, 
S.~Peng~Oh$^1$\thanks{peng@physics.ucsb.edu},
Frederick~B.~Davies$^2$, and 
Steven~R.~Furlanetto$^{2}$\\
$^1$University of California Santa Barbara, Department of Physics, Santa Barbara, CA 93106, USA\\
$^2$University of California Los Angeles, Department of Physics and Astronomy, Los Angeles, CA 90095, USA\\
}

\maketitle

\begin{abstract}
The ionizing background of cosmic hydrogen is an important probe of the sources and absorbers of ionizing radiation in the post-reionization universe.  Previous studies show that the ionization rate should be very sensitive to changes in the source population: as the emissivity rises, absorbers shrink in size, increasing the ionizing mean free path and, hence, the ionizing background. By contrast, observations of the ionizing background find a very flat evolution from $z \sim 2-5$, before falling precipitously at $z\sim 6$. We resolve this puzzling discrepancy by pointing out that, at $z \sim 2-5$, optically thick absorbers are associated with the same collapsed halos that host ionizing sources. Thus, an increasing abundance of galaxies is compensated for by a corresponding increase in the absorber population, which moderates the instability in the ionizing background. However, by $z \sim 5-6$, gas outside of halos dominates the absorption, the coupling between sources and absorbers is lost, and the ionizing background evolves rapidly. Our halo based model reproduces observations of the ionizing background, its flatness and sudden decline, as well as the redshift evolution of the ionizing mean free path. Our work suggests that, through much of their history, both star formation and photoelectric opacity in the universe track halo growth.
\end{abstract}


\section{Introduction}\label{sec:intro}

The ionizing background in the intergalactic medium (IGM) depends on both the production and absorption rate of photons beyond the Lyman limit.  Because stars dominate this production, the ionizing background is an important probe of the buildup of the star formation rate density and evolution in the ionizing escape fraction.  This is particularly true at high-redshift, where a significant population of galaxies lie below the current UV detection limits and escaping Lyman-limit photons are completely absorbed by intervening gas.  \citet{McQuinn11} emphasized the extreme sensitivity of the background ionization rate, $\Gamma$, to the source ionizing emissivity, $\epsilon$, and found $\Gamma \propto \epsilon^{2\text{--}4.5}$, with the exponent increasing toward higher redshifts.

However, these authors also pointed out that this sensitive dependence implies a puzzling inconsistency between recent observations demonstrating the nearly flat evolution of the ionizing background from $z \sim 2\text{--}5$ \citep{Bolton05, Becker07, CAFG08b, BB13} and a rapidly evolving star formation rate density in the universe over the same redshift interval \citep[e.g.,][]{Bouwens12}.  Only at $z>5$ does the background ionization rate appear to evolve rapidly \citep{Fan06a, BH07, WB11, Calverley11}.

Of course, the ionizing escape fraction, $\fesc$, of galaxies, which modulates the star formation rate density to produce the ionizing emissivity, is highly uncertain at high redshift \citep[e.g.,][]{FL13}.  Thus, recent work has focused on fine-tuning the evolution of $\fesc$ to produce consistency between observations of (1) the column density distribution of absorbers, (2) the star formation rate density, and (3) the background ionization rate \citep[e.g.,][]{HM12, KFG12}.  

We suggest that a more generic solution to this puzzle may lie in recent suggestions, both theoretical and observational, that, in addition to the production rate of ionizing photons, galaxies are also connected to Lyman-limit systems (LLSs), which dominate the absorption of such photons \citep[e.g.][]{Rauch08, Steidel10, Rudie12, Font-Ribera12, RH11, McQuinn11, RS14, CAFG14}.  Thus, an increasing ionizing emissivity associated with a growing abundance of galaxies could be balanced by a corresponding increase in the population of absorbers.  If true, quasar absorption lines in general, and the ionizing background in particular, could be useful probes, not only of the star formation in galaxies, but also of the gas in and around them.

In this paper, we test the hypothesis that a link between LLSs and galaxies can explain the flat evolution of $\Gamma(z)$ observed by \citet{BB13}.  We develop simple, semi-analytic prescriptions to describe the distribution and ionization state of gas in dark matter halos as well as the galaxies hosted in the same structures.  We then compute the resulting background ionization rate and its evolution.  Moreover, we show that low-overdensity gas outside halos must contribute to absorption at $z\sim5$, which naturally decouples sources from absorbers and enables the observed drop in $\Gamma$. 

In \S\ref{sec:insight}, we begin by developing analytic insight into the dependence of $\Gamma$ on the ionizing emissivity.  We then describe our semi-analytic models for absorbing gas and the production of ionizing photons associated with dark matter halos in \S\ref{sec:sam}.  In \S\ref{sec:results}, we present our results for the evolution of $\Gamma$, comparing them to recent observations, and discuss the sensitivity of these results to model assumptions in \S\ref{sec:assump}.  Finally, we conclude in \S\ref{sec:conc}.

Throughout this work, we assume a $\Lambda$CDM cosmology with ($h$, $\Omega_{\rm m}$, $\Omega_{\Lambda}$, $\Omega_{\rm b}$, $\sigma_{8}) = (0.7$, 0.28, 0.72, 0.046, 0.82).

\section{Analytic Insight}\label{sec:insight}

To derive physical insight, we begin with analytic scaling arguments similar to those in \citet{McQuinn11}.  In the \citet{MiraldaEscude00} model of the intergalactic medium, gas above a critical overdensity, $\Di$, is assumed to be completely neutral, while less dense material is completely ionized.  This gas, with mean free path $\lambda$, then filters the emission from sources, with ionizing emissivity $\epsilon$, to produce the background ionization rate:
\begin{equation}\label{eq:G}
\Gamma \propto \epsilon\,\lambda.
\end{equation}
Here, the mean free path is related to the properties of a population of identical absorbers by 
\begin{equation}\label{eq:lambda}
\lambda \propto (\na\,\sa)^{-1}, 
\end{equation}
where $\na$ is the number density of absorbers and $\sa \propto \ra^2$ is the absorber cross-section with $\ra$ the typical inverse-Stromgren radius of an absorber setting the size of the optically thick core.  

Let us approximate LLSs as spherically symmetric absorbers, each with density profile $n(r) \propto r^{-\alpha}$, illuminated by a meta-galactic ionizing background.  Outside the core, we assume that the ionization rate equals the recombination rate in the optically thin limit,
\begin{equation}\label{eq:bal}
\Gamma\,\nHI \propto \nH^2,
\end{equation}
where $\nH$ and $\nHI$ are the number densities of the total and neutral gas, respectively.  The size $\ra$ is set by assuming a fixed value of the ionizing optical depth for the gas to become optically thick,
\begin{equation}\label{eq:tau}
\tau = \int_{\ra}^{\infty} \!\! n_{\HI}\,\sigma_{\HI}\,dr \propto \frac{\nH^2(\ra)\,\ra}{\Gamma}\propto {\rm constant},
\end{equation}
where $\sigma_{\HI}$ is the hydrogen ionization cross-section at $912\,{\rm \AA}$.  Substituting for the density, we find 
\begin{equation}
\ra \propto \Gamma^{1/(1-2\,\alpha)},
\end{equation}
which relates the core radius of an absorber to the intensity of the ionizing background.  
Combining with equations~\ref{eq:G} and~\ref{eq:lambda} yields
\begin{equation}\label{eq:emissivity}
\frac{\epsilon}{\na} \propto \ra^{3-2\,\alpha}
\end{equation}
and
\begin{equation}\label{eq:G1}
\Gamma \propto \left(\frac{\epsilon}{\na}\right)^{(2\,\alpha-1)/(2\,\alpha-3)} \propto \left(\frac{\epsilon}{\na}\right)^{\theta},
\end{equation}
where $\theta \equiv \dd\ln{\Gamma}/d\ln{\epsilon}=(2\,\alpha-1)/(2\,\alpha-3)$.  If the density profile of absorbers is isothermal with $\alpha=2$ and their abundance, $\na$, is held fixed, then $\theta=3$, and $\Gamma$ will vary sensitively with $\epsilon$.  In essence, as the ionization rate increases in response to an increasing emissivity, the size of absorbers shrinks, leading to a smaller mean free path and an even larger ionization rate.  This is the same basic result found by \citet{McQuinn11}, who derived similarly sensitive scalings ($\theta \approx 2\text{--}4.5$) using the density profiles in numerical simulations, effectively holding $\na$ constant while varying $\epsilon$.  Note that holding $\na$ constant is also equivalent to adopting the ansatz from \citet{MiraldaEscude00} that the typical distance between absorbers is proportional to the volume filling factor of neutral gas to the $-2/3$ power.

To gain further insight, let us describe the emissivity as the product of the number density of sources, $\ns$, the average ionizing luminosity produced per source, $\Ls$, and the ionizing escape fraction $\fesc$.  While the specific star formation rate and resulting luminosity of galaxies evolves with the cosmic accretion rate and decreases with decreasing redshift \citep[e.g.,][]{Dave12, Munoz12, Stark13}, the main driver for the increasing star formation rate density of the universe down to $z\sim2$ is the growing number of sources \citep[e.g.,][]{Trenti10, ML11}.  Recent studies have effectively balanced this evolving $\ns$ with an evolving $\fesc$ \citep{HM12, KFG12}.  However, if $\fesc$ is held fixed, then, although $\Gamma$ is a sensitive function of $\fesc\,\Ls\,\ns/\na$, the key question is how independent are $\ns$ and $\na$.  If the abundance of sources can grow without changing the absorber population, then we retain the \citet{McQuinn11} result.  However, if $\ns \propto \na$ so that the formation of additional sources also adds proportionally more absorbers to the universe, then the background ionization rate will be relatively insensitive to the changing source emissivity.  

That sources and absorbers are closely related is well-known from numerical simulations \citep[e.g.,][]{RS14}.  The difficulty, however, lies in a self-consistent treatment of the source emissivity and the meta-galactic background over a volume large enough to span the ionizing mean free path while resolving the density distribution in halos around sources.  In the following section, we explore the related evolution of absorbers and sources using a semi-analytic treatment.

\section{Semi-Analytic Treatment}\label{sec:sam}

In this section, we present semi-analytic prescriptions for absorbers (\S\ref{sec:sam:abs}) and sources (\S\ref{sec:sam:emit}) to explore the connection between the two in more detail.  In \S\ref{sec:sam:sum}, we give a brief summary of the models with a list of free parameters.

\subsection{Absorber Model}\label{sec:sam:abs}

\subsubsection{Simplified Density Profile and Ionization Fraction}

We assume that neutral gas associated with dark matter halos, rather than diffuse clouds in the IGM, dominates the column density distribution for LLSs \citep[e.g.,][]{RS14}.  We take the gas profile to trace that of the dark matter in the halo.\footnote{This assumption is unlikely to be correct in detail, but our results are robust to variations.  See \S\ref{sec:assump:profile} on the sensitivity of our results to the choice of profile.}  For an NFW profile \citep{NFW97}, the distribution of over-density, $\Delta$, is
\begin{equation}\label{eq:nfw}
\Delta = \frac{\cvir^3\,\Dvir}{3\,A(\cvir)}\,\left[\frac{1}{(r/\rs)\,(1+r/\rs)^2}\right],
\end{equation}
where $A(\cvir)\equiv \ln(1+\cvir)-\cvir/(1+\cvir)$, $\cvir \equiv \rvir/\rs$ is the halo concentration parameter, and the non-linear critical overdensity enclosed within the virial radius, $\rvir$, required for spherical collapse \citep[e.g.,][]{BL01} is $\Dvir\approx18\,\pi^2$ in the matter-dominated epoch but evolves at low-redshift when dark energy becomes important.

\begin{figure}
\begin{center}
\includegraphics[width=\columnwidth,trim=10 40 20 0,clip]{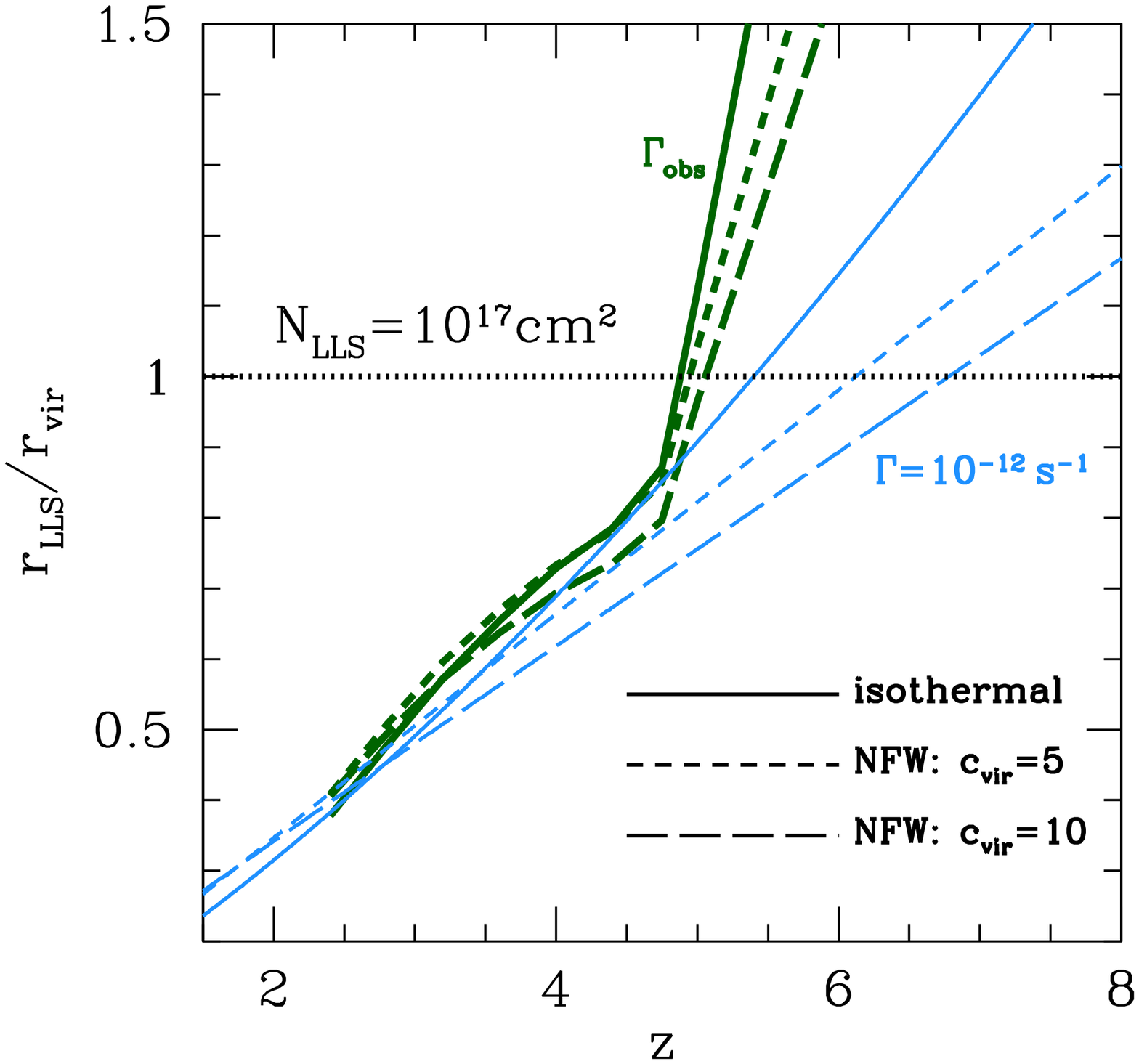}
\caption{\label{fig:rLLS} 
The radius, $\rLLS$, at which the local gas overdensity exceeds the critical value, $\Di$, corresponding to LLSs with $N_{\rm HI} \sim 10^{17}\,\cm^2$ as a function of redshift.  Thin (blue) lines assume a fixed value of $\Gamma=10^{-12}\,\s^{-1}$, while thick (green) lines adopt the \citet{BB13} mean values at $z=2.4$--$4.75$ and an average of measurements from \citet{WB11} and \citet{Calverley11} at $z=5$ and $6$.  We show results for an isothermal halo profile (solid) and NFW profiles with $\cvir=5$ (short-dashed) and 10 (long-dashed).  For reference, $\cvir\approx 3.3$ for $10^{9}\,\msun$ at $z=4$ in the \citet{DM14} model using our cosmology.  The horizontal, dotted line marks $r_{\rm LLS}=r_{\rm vir}$ and indicates that gas outside halos begins to determine the mean free path sometime between redshifts 5 and 6.
}
\end{center}
\end{figure} 

However, in the early universe, the background density outside halos was sufficiently high to dominate the absorption of ionizing photons.  We can estimate the redshift at which this occurs by asking when the over-density at the virial radius of halos exceeds that associated with LLSs, which, in turn, we compute by assuming the \citet{Schaye01} description of identical absorbers with sizes given by the Jeans scale\footnote{The Jeans scale here is the one at which the free-fall and sound-crossing times are equal.  This includes a primary contribution from the dark matter and does not assume that the gas is self-gravitating.} in the optically thin limit of ionization equilibrium \citep[see also][]{FO05}:
\begin{equation}\label{eq:Di}
\begin{split}
\Di \approx 45 \, & \left(\frac{\Gamma}{10^{-12}\,{\rm s^{-1}}}\right)^{2/3}\,\left(\frac{1+z}{7}\right)^{-3} \\
& \times \left(\frac{\NLLS}{10^{17}\,{\rm cm^{-2}}}\right)^{2/3}\,\left(\frac{T}{10^{4}\,{\rm K}}\right)^{0.17},
\end{split}
\end{equation}
where $\NLLS$ is the neutral column density of a LLS.  Since, at fixed $\Gamma$, gas self-shields at fixed physical density, the associated overdensity falls with increasing redshift as reflected in equation~\ref{eq:Di}.

This optically thin ansatz \citep{Schaye03}, which we adopt to drive physical intuition, has been show to successfully find the transition to self-shielding (equations~\ref{eq:xHI}--\ref{eq:nss}) in radiative transfer simulations \citep[e.g.,][see Fig. 5]{McQuinn11}, which is the object of our Figure~\ref{fig:rLLS}.  It is not meant to be used for very optically thick systems. 

Combining equation~\ref{eq:Di} with the density distribution in equation~\ref{eq:nfw} and defining the radius at which $\Delta(\rLLS)=\Di$ to be $\rLLS = \xLLS\,\rvir$, we find
\begin{equation}
\xLLS= \frac{1}{3\,\cvir}\, \left(\tilde{D}+\frac{1}{\tilde{D}}-2\right),\nonumber
\end{equation}
\begin{equation}\label{eq:xnfw}
\tilde{D}=\left[(3/2)\, (81\,D^2+4\,D)^{1/2}+4 D+1\right]^{1/3},
\end{equation}
where $D\equiv(\cvir^3/[3\,A(\cvir)])\,(\Dvir/\Di)$.
The corresponding value for an isothermal profile is
\begin{equation}\label{eq:xIT}
\xLLS^{\rm isothermal}= \left(\frac{\Dvir}{3\,\Di}\right)^{1/2},
\end{equation}
where $\xLLS^{\rm isothermal}=1$ at $z\approx 5.4$ for $\Gamma=10^{-12}\,\s^{-1}$ and $N_{\rm HI} \sim 10^{17}\,\cm^2$.  Note that, at fixed values of $\cvir$ or for an isothermal profile, $\xLLS$ is independent of halo mass.  Indeed, the transition redshift would not change if we adopted the splashback radius---defined as the apocenter of particles on their first orbit after being accreted---for the natural halo boundary, as recently suggested by \citet{More15} and instead of the virial radius.  This is because, at $z=5$, $\rsp$ is within $20\%$ of $\rvir$ for $10^{8}\,\msun$ halos and within $3\%$ for $10^{11}\,\msun$ halos.

Figure \ref{fig:rLLS} compares $\rLLS$ as a function of redshift for different choices of the density profile.  At low redshifts, where $\Di$ is large, sufficiently neutral gas resides only in the inner regions of the halos where $r<\rs$.  Here, the resulting values of $\rLLS$ are very similar among different profile choices.  However, at larger redshifts, the value of $\Di$ becomes only quasi-linear, and $\rLLS$ exceeds the virial radius.  At this point, outside the virial radius, $\rLLS$ depends more strongly on the assumed shape of the density profile.  The transition between the two regimes occurs sometime between redshifts 5 and 6 and is more rapid if we assume the observations of $\Gamma(z)$ as given (thick curves in the figure) because the decreasing ionization rate after $z \sim 5$ contributes to additional growth in $\rLLS$.

The above analysis supposes that the Lyman continuum opacity of the IGM is dominated by absorbers with column densities above $\NLLS$.  This appears to be the case at $z\sim2$ \citep{HM12}, though the extrapolation to higher redshifts is less clear.  In equation~\ref{eq:dteff}, we will relax this simplifying assumption and integrate over the entire column density distribution function.

\subsubsection{Detailed Treatment}\label{sec:sam:abs:det}

For a more detailed treatment of the halo density profile, we adopt the \citet{DM14} fitting model for the NFW concentration parameter as a function of halo mass and redshift to describe the profile in the inner parts of the halo.  We additionally assume that the halo transitions to a flatter profile in its outskirts to match onto the mean IGM density at very large radii.  Quantifying this transition has been the subject of much recent work in the literature \citep{Prada06, HW08, Cuesta08, Tavio08, OH11, BK11, DK14, More15}.  We adopt an outer density profile derived from a suite of N-body simulations of dark matter by \citet[][see \S\ref{sec:app:dk} for the details of our implementation]{DK14}.  

To examine specifically the effect of gas within the halo on the flatness of the ionizing background, we additionally truncate the density profile at radii beyond the splashback radius, $\rsp$.  We adopt the fitting formulae derived from numerical simulations by \citet{More15} and set $\Delta(r>\rsp)=0$.

With the halo profile specified, we can use our halo-based absorber model to determine the probability distribution function (PDF) of overdensities $\Delta>\Di$ by calculating the fractional volume occupied by each overdensity around a halo and integrating over the halo mass function, $\dd n/\dd M$, which we take to be Sheth-Tormen mass function \citep{ST99, SMT01}, from $\mabs$ to infinity.  $\mabs$ is the minimum halo mass capable of hosting absorbing gas and effectively controls the number of absorbers.  Below $\mabs$, halos cannot retain their gas and infall of new gas is suppressed.  For a given value of $\mabs$, the PDF of overdensity is
\begin{equation}\label{eq:pdf}
\frac{\dd P_{\rm V}(\Delta_0,z)}{\dd \log\Delta}=\int_{\mabs}^{\infty} \!\! 4\,\ln{10}\,\pi\,r_{0}^3\,(1+z)^3\,\left[\frac{\dd \ln\Delta}{\dd \ln{r}}\right]_{r_0}^{-1}\,\frac{\dd n}{\dd M}\,dM,
\end{equation}
where $r_0$ is the radius at which the overdensity of a halo is $\Delta_0$; $\dd\ln\Delta/\dd\ln r$ is evaluated at $r_0$; and $r_0$, $\dd\ln\Delta/\dd\ln r$, and $\dd n/\dd M$ are each functions of $M$ and $z$.  
 
\begin{figure}
\begin{center}
\includegraphics[width=\columnwidth,trim=15 20 25 0,clip]{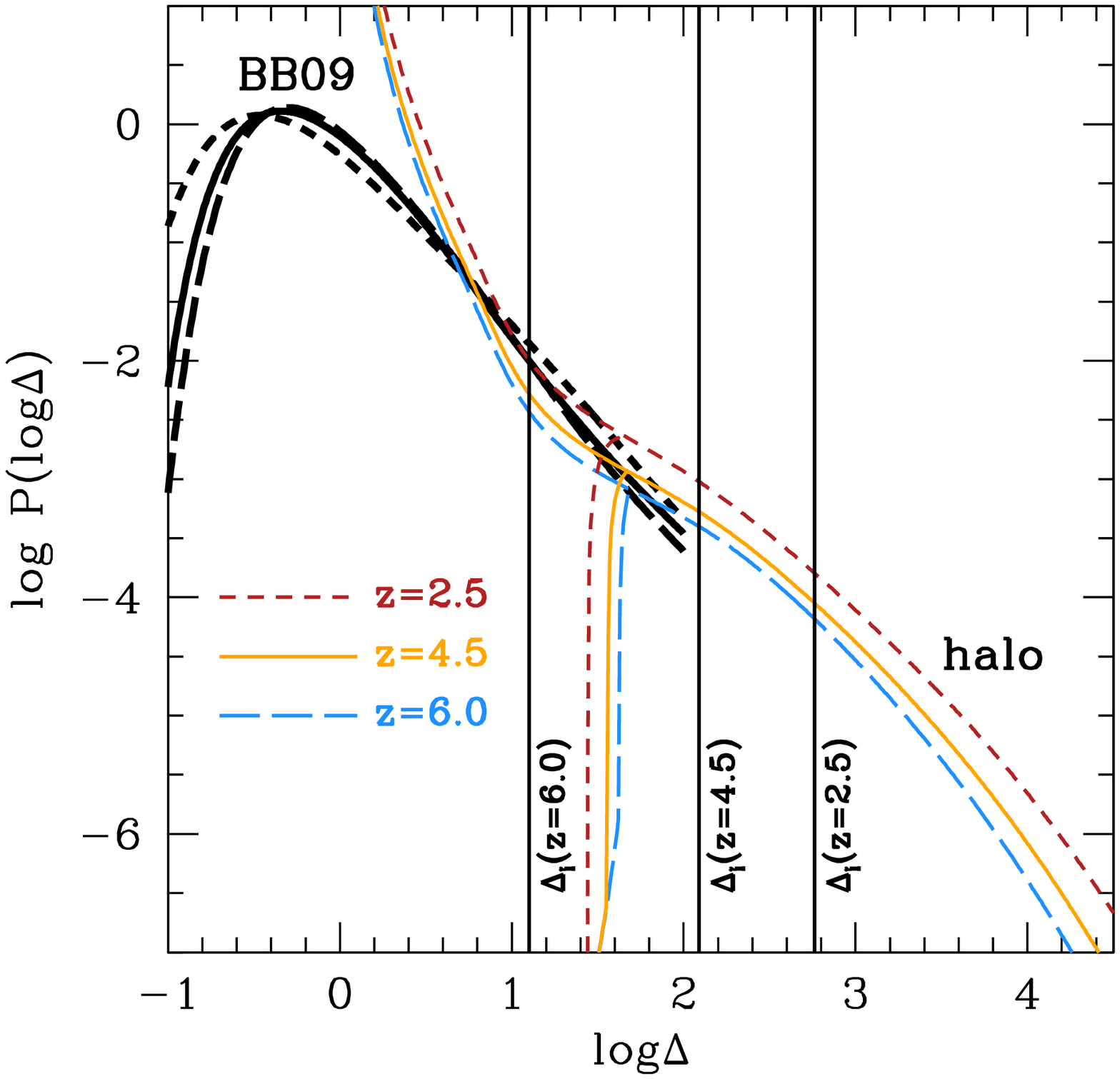}
\caption{\label{fig:pdf} 
The PDF of overdensity, where $P(\log\Delta)\equiv \dd P_{\rm V}/\dd\log\Delta$.  Thick and thin curves show results from the numerical simulations of \citet{BB09} and from equation~\ref{eq:pdf} with $\mabs=\mf$, respectively, at redshifts $z=2.5$ (short-dashed), 4.5 (solid), and 6.0 (long-dashed).  The set of results showing a sharp drop in probability density at $\log\Delta\lesssim1.6$ truncate the halo density profile at radii larger than the splashback radius, $\rsp$.  Vertical lines denote the overdensity corresponding to LLS with $\NHI=10^{17}\,\cm^{-2}$ from equation~\ref{eq:Di} assuming $\log \Gamma/10^{-12}\,\s^{-1}= 0$, 0, and $-0.835$ at $z=2.5$, 4.5, and 6.0, respectively, consistent with observations.
}
\end{center}
\end{figure}

In Figure~\ref{fig:pdf}, we plot our model results for the gas around halos at $z=2.5$, 4.5, and 6.0.  To compare more realistically to results from numerical simulations in the literature, we set $\mabs$ equal to the filtering mass \citep[][using the updated definition of \citealt{Naoz09}]{Gnedin00}:
\begin{equation}
\mf = 2.2\times10^{10}\,\msun\,\left(\frac{1+z}{5}\right)^{-3/2}\,\tilde{f}(z, \zrei)^{3/2}, \nonumber
\end{equation}
\begin{equation}\label{eq:mf}
\tilde{f}(z, \zrei)=0.3\,\left[1+4\,\left(\frac{1+z}{1+\zrei}\right)^{2.5}-5\,\left(\frac{1+z}{1+\zrei}\right)^{2}\right],
\end{equation}
where we assume a reionization redshift of $\zrei=9$ throughout this work.  This prescription effectively evaluates the Jeans criterion at the mean density of the universe without accounting for the detailed formation histories of halos \citep{NM14}.  Nevertheless, it serves as a standard test case.  At the two lower redshifts, $\Di$ is in the `halo' regime within the splashback radius captured by our model.\footnote{Note also that our density PDF is unlikely to be accurate for $\Delta \gsim 300$, when the effects of gas cooling, star formation and feedback become particularly important. In the redshift range of interest, $\Delta_{i}$ generally lies below this.}  Between $z=4.5$ and $z=6.0$, cutting off the density profile at $\rsp$ produces a sharp turnover in the PDF toward lower overdensities.  However, without that cutoff, the figure shows that the density profile would rise indefinitely.  This represents a shortcoming in using any average profile to describe halos, even one that matches onto the mean density at large radii like the one in \citet{DK14}: under-dense fluctuations about the mean density will always be missing.  In addition, using the profiles of halos at large radii results in double counting once the density distributions of halos overlap.

However, the density PDF in the diffuse IGM has been probed by numerical simulations \citep[e.g.,][]{MiraldaEscude96, MiraldaEscude00, BB09, McQuinn11}.  Figure~\ref{fig:pdf} compares our halo results to recent \citet{BB09} predictions using the convenient fitting formulae supplied.
At each plotted redshift, the simulated PDF intersects the one built on the halo density profile at the point where our fiducial $\Delta(r>\rsp)=0$ calculation sharply turns over.  This suggests a prescription for constructing a composite PDF using the maximum of the \citet{BB09} and halo results at each overdensity.

In addition to the overdensity, we can also directly compute the column densities of neutral gas associated with halos in our halo-based model.  The physical volume density of neutral gas is $\nHI=\xHI\,\nH$, where $\nH=\Delta\,\rho_{\rm crit}\,\Omega_{\rm b}\,(1-\Yhe)/m_{\rm p}$, the helium fraction is $Y_{\rm He}=0.24$, the neutral gas fraction $\xHI$ is given by photoionization equilibrium\footnote{Studies that have included collisional ionization have found only a modest effect---at most $50\%$ and only over a narrow range of densities \citep[][Figs. 4 and 6]{Rahmati13a}---though the details depend on the feedback assumptions.}:
\begin{equation}\label{eq:xHI}
\xHI\,\Gloc=\alpha_{\rm rec}\,\nH\,(1-\xHI)^2,
\end{equation}
$\alpha_{\rm rec}$ is the recombination coefficient, and $\Gloc$ is the ionization rate in gas with hydrogen density $\nH$ and subject to a background ionization rate $\Gamma$.  We compute $\Gloc$ using the prescription derived by \citet{Rahmati13a} from numerical simulations that include radiative transfer and the effects of self-shielding:
\begin{equation}\label{eq:Gloc}
\frac{\Gloc}{\Gamma}=0.98\,\left[1+\left(\frac{\nH}{\nss}\right)^{1.64}\right]^{-2.28}+0.02\,\left[1+\frac{\nH}{\nss}\right]^{-0.84},
\end{equation}
where
\begin{equation}\label{eq:nss}
\nss \approx 6.73 \times 10^{-3}\,\cm^{-3}\,\left(\frac{T}{10^{4}\,\K}\right)^{0.17}\,\left(\frac{\Gamma}{10^{-12}\,\s^{-1}}\right)^{2/3}
\end{equation}
is the number density at which the gas begins to self-shield\footnote{The result in equation~\ref{eq:nss} is analogous to that in equation~\ref{eq:Di}; both give the correspondence between density and ionization rate.  Indeed, the two agree to within a factor of 2.}, and $T$ is the temperature of the IGM.  We set a constant temperature of $T=10^{4}\,{\rm K}$ throughout this work, but note that equation \ref{eq:nss} depends only weakly on this choice.  We further take $\alpha_{\rm rec}$ to be the value for Case A recombinations, $\alpha_{\rm A}\approx 4.2\times10^{-13}\,(T/10^4\,K)^{-0.76}$, since equation~\ref{eq:Gloc} automatically includes the effect of recombination radiation \citep{Rahmati13a}.  Finally, we emphasize that, though we adopted the most accurate prescriptions available to implement in our model, the details of these prescriptions are not critical to our results (see \S\ref{sec:assump:profile}).

We can then obtain the column density distribution of neutral gas by using equations~\ref{eq:xHI},~\ref{eq:Gloc}, and~\ref{eq:nss} and computing the fractional projected area around each halo occupied by lines of sight with a given column density.  After integrating over $\dd n/\dd M$, the canonical distribution function is
\begin{equation}\label{eq:cddf}
f(N_{\HI, 0},z)=\int_{\mabs}^{\infty} \!\! \frac{2\,\pi\,c}{H_0}\,\frac{b^2_{0}(N_{\HI, 0})}{N_{\HI,0}}\,\left[\frac{\dd \ln\NHI}{\dd \ln{b}}\right]_{b_{0}}^{-1}\,\frac{\dd n}{\dd M}\,dM,
\end{equation}
where
\begin{equation}
f(N_{\rm HI,0},z)\equiv  \frac{\deriv^{2}N}{\deriv N_{\rm HI} \deriv z} \frac{H(z)}{H_{0}} \frac{1}{(1+z)^{2}},
\end{equation}
and the impact parameter $b_0$ corresponding to $N_{\HI,0}$ is given by \citep[see also][]{MI90}
\begin{equation}\label{eq:b}
N_{\HI,0}=2\,b_{0}\,\int_{1}^{\infty}\!\! n_{\HI}(b_{0}\,y,M,z)\,\frac{y\,dy}{\left(y^2-1\right)^{1/2}}.
\end{equation}
Note that we evaluate $\dd \ln\NHI/\dd \ln{b}$ at $b_0$ and that $b_0$, $\dd \ln\NHI/\dd \ln{b}$, and $\dd n/\dd M$ are each functions of both $M$ and $z$.

\begin{figure}
\begin{center}
\includegraphics[width=\columnwidth,trim=5 20 10 0,clip]{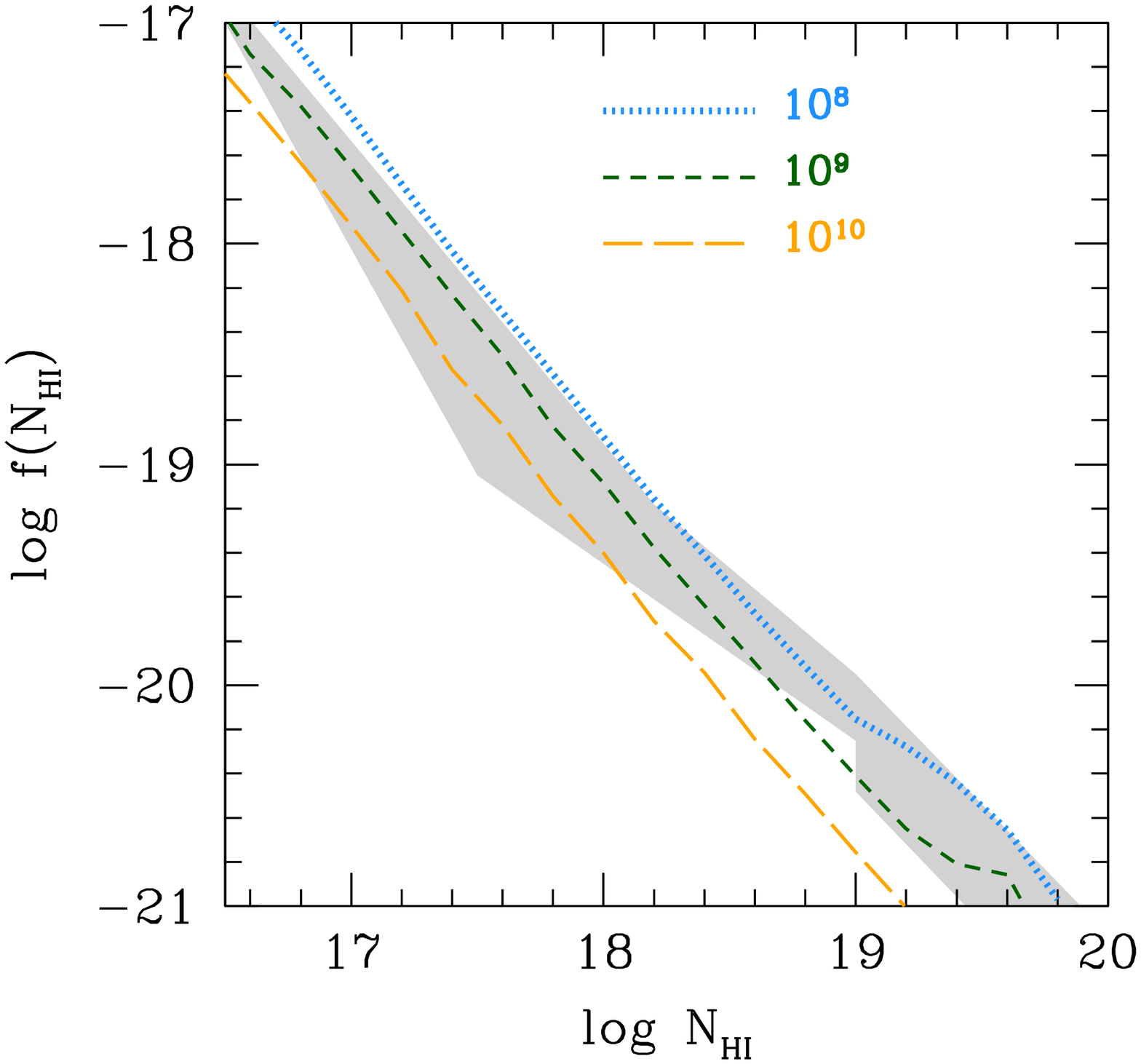}
\caption{\label{fig:cddf} 
The column density distribution function at $z=4$.  Dotted (blue), short-dashed (green), and long-dashed (orange) curves show the computation from equation~\ref{eq:cddf} for $\log\mabs/\msun=8$, $9$, and $10$, respectively, using the source model in \S\ref{sec:sam:emit} with $\log\mabs/\msun=\log\memit/\msun$ and values of $\fesc$ best fit to the \citet{BB13} background ionization rate measurements.  The shaded region denotes the observational compilation in \citet{Prochaska10} and is well-described by our model with $\log\mabs/\msun=9$ at this redshift (see text). 
}
\end{center}
\end{figure}

In Figure \ref{fig:cddf}, we compare the resulting distribution of column densities at $z=4$ for different values of $\mabs$ to the compilation of observations in \citet{Prochaska10}.\footnote{The results in Fig.~\ref{fig:cddf} assume the emitter model in \S\ref{sec:sam:emit} and have been iterated for convergence in $\Gamma$ and fit to the ionizing background observations of \citet{BB13}.}  At this redshift, we find good agreement from $\NHI \sim 10^{16.5}\text{--}10^{20} \, {\rm cm^{2}}$ using $\mabs = 10^{9}\,\msun$, a mass inferred from numerical simulations \citep[e.g.,][and references therein]{NM14} and consistent with the filtering mass at $z=4$ and the minimum mass required for sources as determined from observations of the UV galaxy luminosity function \citep{ML11}.  Moreover, our column density distribution is also similar to the models published in \citet{Rahmati13a}, where self-shielding produces only a modest deviation from a power-law over this range of column densities \citep[see also][]{SM14}.  The steep slope ensures that most of the opacity arises at the Lyman limit and that $\mfp$ is primarily sensitive to the total abundance of absorbers, which is effectively set by $\mabs$, rather than the details of their column density distribution.  Thus, while the effects of self-shielding may be starker at still higher column densities, we stress that these differences have little effect on the mean free path.

The column density distribution directly yields the mean free path of the IGM, which we can evaluate at the Lyman limit: 
\begin{equation}\label{eq:mfp}
\mfp(z) = \frac{\dd l}{\dd z}\,\left[\frac{\dd \tau_{\rm eff}(\nu_{912},z)}{\dd z}\right]^{-1},
\end{equation}
where
\begin{equation}\label{eq:dteff}
\frac{\dd \tau_{\rm eff}(\nu,z)}{\dd z}=\int_0^{\infty}\!\! \dd \NHI\,\frac{\dd^2 N}{\dd \NHI\, \dd z}\,\left[1-{\rm e}^{-\NHI\,\sigma_{\HI}(\nu)}\right],
\end{equation}
$\dd l/\dd z=c\,H^{-1}(z)\,(1+z)^{-1}$ is the proper distance per redshift interval for an evolving Hubble parameter $H(z)=H_{0}\,\sqrt{\Omega_{\rm m}\,(1+z)^3+\Omega_{\Lambda}}$, and where $\sigma_{\HI}(\nu)=\sigma_{912}\,(\nu/\nu_{912})^{-3}$ is the ionization cross-section of neutral hydrogen with a value of $\sigma_{912}\approx6.3\times10^{-18}\,\cm^{2}$ at $\nu_{912}=3.29 \times 10^{15}\,\s^{-1}$, the frequency corresponding to the Lyman limit.  Because of the exponential term in equation~\ref{eq:dteff}, column densities greater than $\sim 1/\sigma_{912}\approx 1.6 \times 10^{17}\,\cm^{-2}$---that is, approximately the value corresponding to LLSs---will dominate the integral.  However, note that we do include the contribution from optically thin absorbers at lower column densities.

\subsection{Emitter Model}\label{sec:sam:emit}

\subsubsection{Galaxies}\label{sec:sam:emit:gal}

We adopt a simple model for the evolving ionizing emissivity of the universe resulting from star formation in galaxies.  We compute the star formation rate within a dark matter halo as a function of its mass and redshift in a way specifically designed to reproduce observations of the UV luminosity function of Lyman-break galaxies and its evolution \citep{Munoz12}:\footnote{For convenience, we ignore the scatter in star formation rate at fixed halo mass, which predominantly affects only the brightest end of the luminosity function.}
\begin{equation}\label{eq:sfr}
\sfr=\frac{\macc}{1+\eta_{\rm w}},
\end{equation}
where \citep{McBride09}
\begin{equation}\label{eq:acc}
\macc\approx 3\,{\rm \msun/yr}\,\left(\frac{\mhalo}{10^{10}\,\msun}\right)^{1.127}\,\left(\frac{1+z}{7}\right)^{2.5} 
\end{equation}
at high redshift and $\eta_{\rm w}\approx (400\,\kms)/\sigma$.  The halo velocity dispersion, $\sigma$, is a function of halo mass and redshift \citep{BL01}:
\begin{equation}
\begin{split}
\sigma=46\,\kms\,&\left(\frac{M_{\rm halo}}{10^{10}\,\msun}\right)^{1/3}\,\left(\frac{1+z}{7}\right)^{1/2} \\
&\times\left[\frac{\Omega_{\rm m}\,h^2/\Omega_{\rm m}(z)}{0.137}\,\frac{\Delta_{\rm c}}{18\,\pi^2}\right]^{1/6}.
\end{split}
\end{equation}
For a $10^{10}\,(10^{12})\,\msun$ halo at $z=3$, $\sigma \approx 35\,(160)\,\kms$, $1+\eta_{\rm w}\approx 13\,(3.5)$, and the average star formation rate is about $0.06\,(40)\,\msun/\yr$.  

\begin{figure}
\begin{center}
\includegraphics[width=\columnwidth,trim=5 20 10 0,clip]{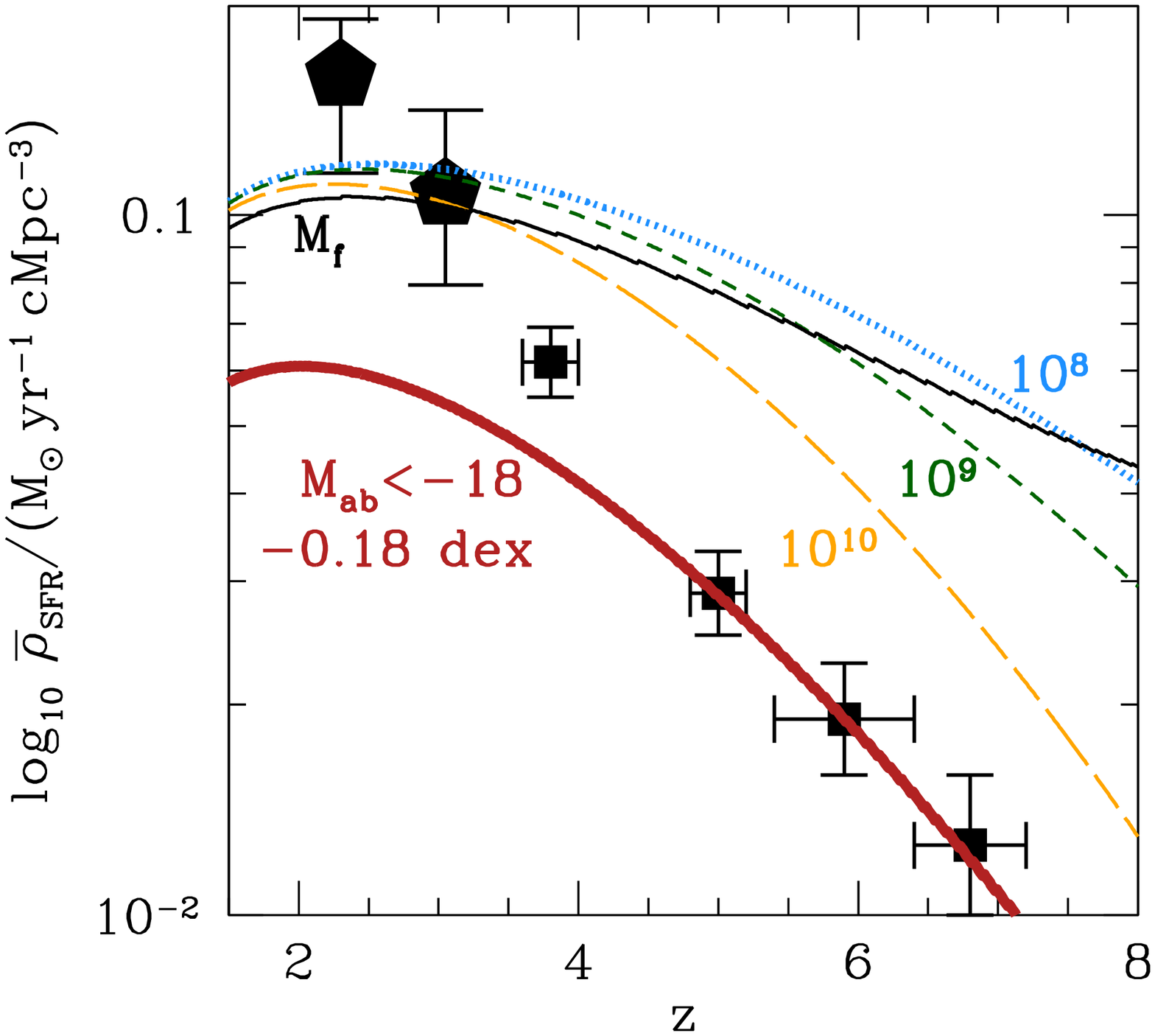}
\caption{\label{fig:sfrd} 
The evolving star formation rate density of the universe.  Solid (black), dotted (blue), short-dashed (green), and long-dashed (orange) curves show results from our emitter model (equation~\ref{eq:sfrd}) for $\log\memit/\msun=\mf$, $8$, $9$, and $10$, respectively.  Pentagons show observational results from \citet{RS09} with an applied correction for dust and undetected sources.  Squares denote measurements by \citet{Bouwens07} and \citet{Bouwens12} only for galaxies brighter than a rest-frame UV absolute magnitude of -18 with no dust correction; the thick, solid (red) curve shows our model adopting this same limiting magnitude and removing a dust correction of $0.18$ dex (see text).
}
\end{center}
\end{figure}

To obtain the ionizing emissivity resulting from this galaxy model, we first compute the comoving star formation rate density of the universe by integrating equation~\ref{eq:sfr} over the halo mass function,
\begin{equation}\label{eq:sfrd}
\bar{\rho}_{\rm SFR}(z) = \int_{\memit}^{\infty}\!\! \sfr(M,z)\,\frac{\dd n}{\dd M}(M,z)\,dM.
\end{equation}
We plot the results in Figure~\ref{fig:sfrd} as a function of redshift for different values of $\memit$ to demonstrate the changing evolution of $\bar{\rho}_{\rm SFR}$ with minimum mass; as $\memit$ increases, it moves into the tail of the mass function, and the abundance changes more rapidly.  The figure further shows the consistency between our model and observationally-based measurements from \citet{RS09} at $z\sim 2$--$3$ that include corrections for both dust extinction and faint sources below the detection limit.  At higher redshift, the \citet{Bouwens07} and \citet{Bouwens12} data reflect only detected galaxies and do not include a contribution from fainter objects, which could dominate the star formation rate density \citep[e.g.,][]{WL06, HB06, ML11}.  Therefore, we expect these points to be only lower limits to the star formation rate density at high redshift.  For comparison, we also show the results from our model if we remove a dust correction to the luminosity of $0.18$ dex, consistent with the determination by \citet{Bouwens07} at $z=6$.  The agreement between this result and the observed points at high redshift is not coincidental; recall that our model is based on a fitting to the observed luminosity function.  Correcting for undetected sources flattens the comoving evolution of the source population but is neither sufficient to explain the flat evolution in $\Gamma$ given the extreme sensitivity between $\Gamma$ and $\epsilon$ derived by \citet{McQuinn11} nor does it solve the issue of fine-tuning.

From the star formation rate density, we obtain the comoving Lyman-limit emissivity, 
\begin{equation}\label{eq:egal}
\epsilon_{912}^{\rm gal}(z)=\fesc\,\lion\,\bar{\rho}_{\rm SFR}(z),
\end{equation}
by assuming values of the ionizing luminosity at $912\,\AA$, $\lion$, produced per star formation rate and of the ionizing escape fraction, $\fesc$, that are independent of both mass and redshift.  While the former depends on the properties of the stellar population, the latter is still more uncertain with theoretical predictions generally conflicted about its dependence on mass and redshift \citep[see][and references therein]{FL13}.  Because our model depends only on the product of the two quantities, we will, in practice, set $\lion=2.7\times10^{27}\,\erg\,\s^{-1}\,\Hz^{-1}\,\msun^{-1}\,\yr$ and leave $\fesc$ as a free parameter.  Note that our resulting escape fractions will change for different assumed values of $\lion$.
We then specify the source spectrum such that $\epsilon_{\nu}(z)=\eion(z)\,(\nu/\nu_{912})^{-\alpha_{912}}$ and assume $\alpha_{912}=2.0$ to be consistent with the method in \citet[][see the discussion in \S5.1 of their paper]{BB13}.  While in principle $\alpha_{912}$ and $\fesc\,\lion$ may vary with redshift and/or halo mass, we assume constant values here to demonstrate that such variation over the redshift range from $z\sim2\text{--}5$ is unnecessary to produce the flat evolution in the background ionization rate, in contrast with studies that invoke redshift-dependent escape fractions \citep[e.g.,][]{HM12, KFG12, FL13}, which necessarily involve some fine-tuning.  Of course, $\fesc$ may additionally vary at still higher redshifts to facilitate cosmic reionization.

\subsubsection{AGN}\label{sec:sam:emit:agn}

In addition to galaxies, AGN may be important sources of ionizing radiation at the redshifts of interest.  Combining optical, UV, and X-ray observations from wide-field samples, \citet{Cowie09} measure the evolution of the comoving AGN ionizing emissivity, peaking around $z=2.2$ at a value of roughly $1.9\times10^{24}\,{\rm erg\,s^{-1}\,Hz^{-1}\,\mpc^{-3}}$.  A competing estimate appears in \citet{HM12}, where the authors adopt the evolving quasar emissivity from \citet{Hopkins07} who integrate the observed luminosity function down to $-27$ magnitudes in the rest-frame B-band and assume a conversion factor from the B-band to the Lyman-limit based on composite spectra that is independent of luminosity and redshift.  The resulting \citet{HM12} fitting formula comoving emissivity is given by
\begin{equation}\label{eq:eagn}
\epsilon_{912}^{\rm AGN}(z) = \frac{(10^{24.6}\,{\rm erg\,s^{-1}\,Hz^{-1}\,Mpc^{-3}})\,(1+z)^{4.68}\,{\rm e}^{-0.28\,z}}{{\rm e}^{1.77\,z}+26.3}.
\end{equation}
While this calculation has, perhaps, fewer uncertainties than does the determination of the galaxy emissivity, the \citet{Cowie09} and \citet{HM12} estimates differ by a factor of about 4.  Citing the \citet{Cowie09} result, \citet{BB13} argue that AGN make a negligible contribution to the emissivity.  Additionally supporting this conclusion are efforts to directly measure ionizing radiation from galaxies, which find that the galaxy population can provide more than enough ionizing flux to explain the observed background \citep{Nestor13, Mostardi13, Mostardi15}.

Given the uncertainties in whether, when, and by how much quasars contribute to $\eion$, we adopt a flexible model in which the total emissivity is  
\begin{equation}\label{eq:etot}
\eion(z)=\epsilon_{912}^{\rm gal}(z)+\fagn\,\epsilon_{912}^{\rm AGN}(z),
\end{equation}
with $\epsilon_{912}^{\rm gal}$ and $\epsilon_{912}^{\rm AGN}$ given by equations~\ref{eq:egal} and~\ref{eq:eagn}, respectively.  Setting $\fagn=1$ is equivalent to adopting the AGN emissivity from \citet{HM12}, while $\fagn=0.25$ approximately reproduces the observed AGN emissivity from \citet[][see Fig.~\ref{fig:abs}]{Cowie09}.

\subsection{Summary of Method}\label{sec:sam:sum}

We construct a model for absorbers (\S\ref{sec:sam:abs}) in which the gas dominating the ionizing mean free path of the IGM is associated with dark matter halos above a minimum mass $\mabs$ and traces an NFW density profile with concentration given by \citet{DM14} in the inner parts of the halo and transitions to a flatter profile prescribed by \citet{DK14} before truncating at the splashback radius, $\rsp$.  Halos above a minimum mass $\memit$ also host galaxies, which we assume to be sources of ionizing radiation (\S\ref{sec:sam:emit}), specifying the star formation rate as a function of halo mass and redshift to reproduce observations of the galaxy UV luminosity function and setting constant values for both the ionizing luminosity produced per star formation rate, $\lion$, and the ionizing escape fraction, $\fesc$.  We approximate the contribution to the ionizing emissivity from AGN by including an additional component which we assume to be a factor $\fagn$ times the \citet{HM12} level.  The combined absorber$+$source semi-analytic model, thus, has four free parameters: $\mabs$, $\memit$, the product $\fesc\,\lion$, and $\fagn$.\footnote{In principle, we could also vary $\alpha_{912}$, the spectral slope of the emitters.  However, within reasonable limits this has very little impact on our results.}  However, in practice, we only allow $\fesc$ to vary and fix the remaining parameters to well-motivated values (see Table~\ref{tab:results}).  In particular, we typically set $\mabs=\memit$ to highlight the relationship between sources and absorbers (while further reducing the number of free parameters), but note that the flat evolution that we find in $\Gamma(z)$ does not depend on a precise equivalence between these two minimum masses (see \S\ref{sec:assump:equiv}).  

\section{Results}\label{sec:results}

In \S\ref{sec:results:flatness} we present results for our model of coupled absorbing gas and ionizing sources inside halos and show that it reproduces the observed flat evolution in $\Gamma$ from $z\sim2$--5.  Then, in \S\ref{sec:results:dropoff}, we show that a contribution to the absorption from uncorrelated, low-overdensity outside halos gas is responsible for the rapid evolution in $\Gamma$ observed at even higher redshifts by incorporating this gas into our model via a composite overdensity PDF.

\subsection{The Flatness of $\Gamma$ from $z\approx 2$--$5$}\label{sec:results:flatness}

Combining our absorber and source models for halos out to the splashback radius, we compute the ionizing background as
\begin{equation}\label{eq:Gamma}
\Gamma(z_0) = 4\,\pi\,\int_{\nu_{912}}^{\infty}\! \frac{\dd \nu_0}{h_{\rm p}\,\nu_0}\,J_{\nu_0}(z_0)\,\sigma_{\HI}(\nu_0),
\end{equation}
where $h_{\rm p}$ is the Planck constant,
\begin{equation}\label{eq:Jnu}
J_{\nu_0}(z_0) = \frac{1}{4\,\pi}\,\int_{z_0}^{\infty}\! \dd z\,\frac{\dd l}{\dd z}\left(\frac{1+z_0}{1+z}\right)^3\,\epsilon_{\nu}(z)\,{\rm e}^{-\tau_{\rm eff}(\nu,z,z_0)},
\end{equation}
$\nu=\nu_0\,(1+z)/(1+z_0)$, and
\begin{equation}\label{eq:teff}
\tau_{\rm eff}(\nu,z,z_0)=\int_{z_0}^{z}\! \dd z'  \, \frac{\dd \tau_{\rm eff}(\nu,z')}{\dd z'}
\end{equation}
with $\dd \tau_{\rm eff}/\dd z$ given by equation~\ref{eq:dteff}.
Equation~\ref{eq:Gamma} includes the redshifting effects of cosmological expansion, important at $z\lesssim3$ when the mean free path is comparable to the proper size of the universe.  At higher redshifts, the background ionization rate is simply proportional to the product of the emissivity and the mean free path as in equation~\ref{eq:G}.  However, $\Gamma$ itself is also an input into the mean free path where it controls the ionization fraction $\xHI$ and, consequently, the column density distribution.  Therefore, to obtain final values of $\mfp$ and $\Gamma$, we begin with a starting value of $\Gamma=10^{-12}\,\s^{-1}$ at all redshifts and iterate equation~\ref{eq:Gamma} until convergence.  

\begin{figure*}
\centering
\begin{tabular}{cc}
\includegraphics[width=0.48\textwidth,trim=53 65 300 20,clip]{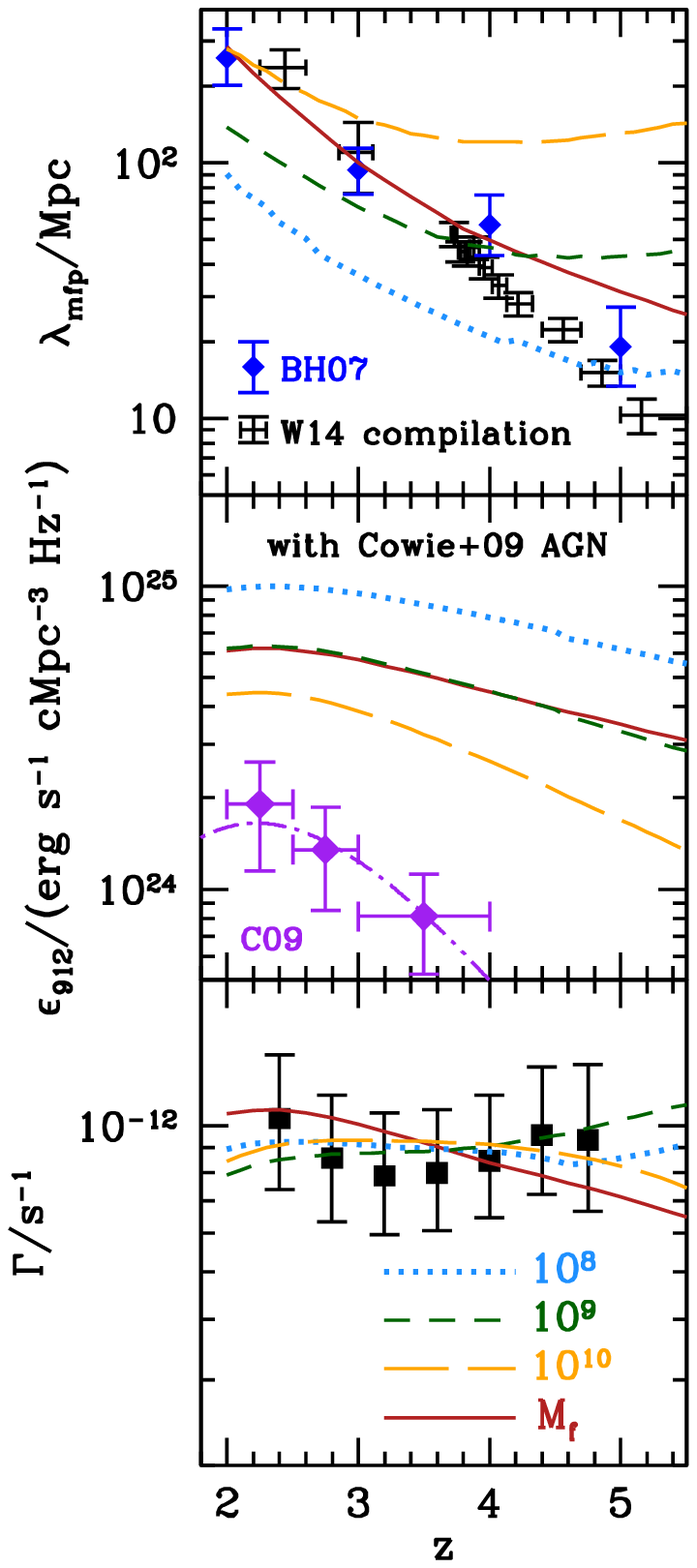} &
\includegraphics[width=0.48\textwidth,trim=48 65 305 20,clip]{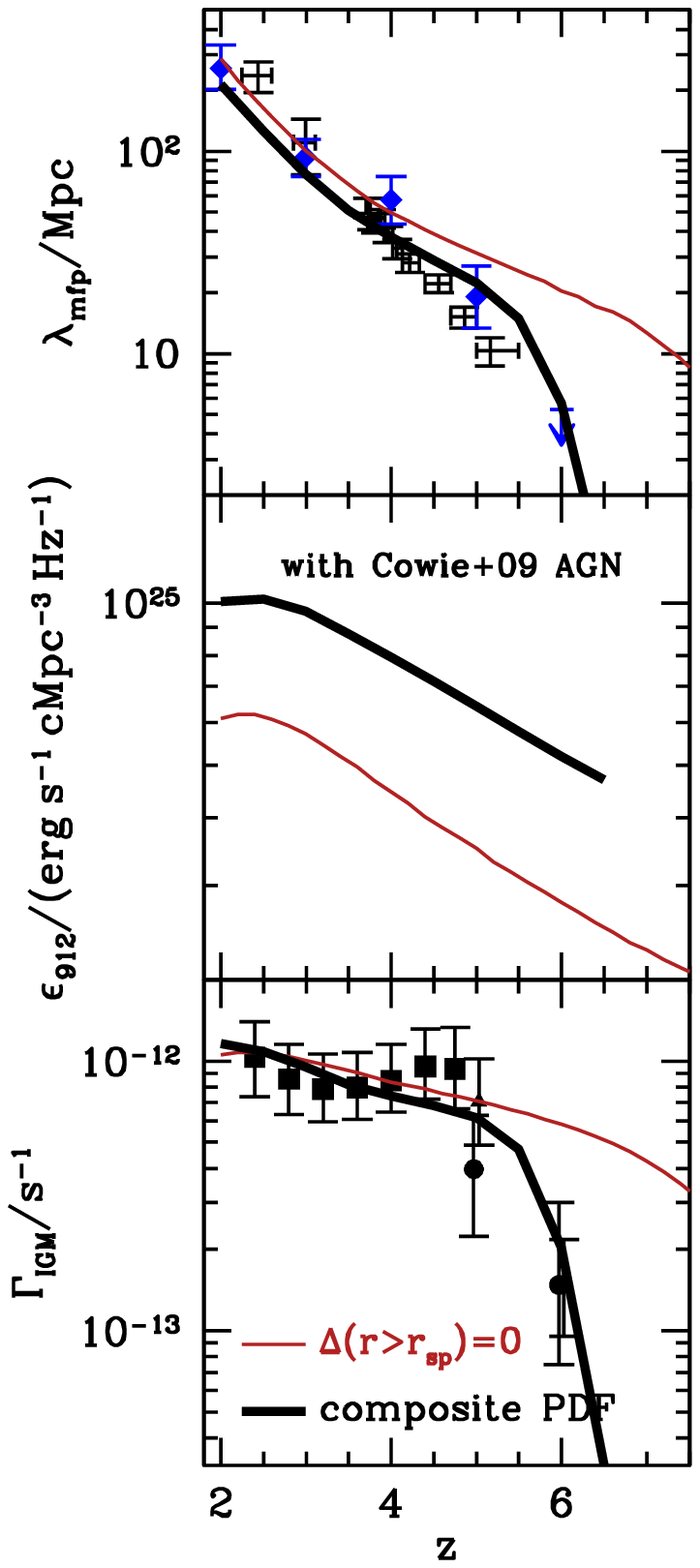}  \\
\end{tabular}
\caption{\label{fig:abs} 
The redshift evolution of the mean free path of the IGM (top), the ionizing emissivity (middle)---both evaluated at the Lyman limit---and the background ionization rate (bottom).  Mean-free-path measurements are taken from \citet{Bolton05} and \citet[][blue diamonds collectively denoted BH07]{BH07} and the compilation in \citet[][black errorbars denoted W14]{Worseck14}.  \citet{BB13} measurements of $\Gamma$ are shown as square (black) points.  In the left-hand column, solid (red), dotted (light-blue), short-dashed (dark green), and long-dashed (orange) curves show the model results best-fit to the \citet{BB13} data for $\log\mabs/\msun=\log\memit/\msun=\mf$ (equation~\ref{eq:mf}), $8$, $9$, and $10$, respectively (see Table~\ref{tab:results}).  All plotted models also assume $\fagn=0.25$, consistent with observations from \citet[][purple diamonds, C09]{Cowie09}.  In the right-hand column, we reproduce the thin, solid (red) curve from the left-hand panels.  Additionally, the thick, solid (black) curve shows the calculation using our composite PDF with equations~\ref{eq:pdf-mfp} and~\ref{eq:Gamma1} and $\log\mabs/\msun=\log\memit/\msun=\mf$.  For comparison at $z=5$ and 6, we have also included the measurements of $\Gamma$ from \citet[][adjusted by \citealt{BB13}, circles]{WB11} and \citet[triangles]{Calverley11}, artificially separated slightly in redshift for clarity.
}
\end{figure*}

We primarily compare our results to the measurements from \citet{BB13} from $z=2.4$--$4.75$.  These authors computed $\Gamma(z)$ by comparing calibrations from numerical simulations with observed IGM optical depths from stacked samples of Sloan Digital Sky Survey quasar absorption spectra \citep{Becker13}.  Their determinations are summarized in Tables 1 and 2 of \citet{BB13}.  Because the resulting values of $\Gamma$ at different redshifts are correlated, in appendix~\ref{sec:app:fit}, we combine their published covariance matrix of statistical errors with their estimated Jeans smoothing and systematic uncertainties to produce a total covariance matrix with which to judge the goodness of fit between our model and the observations.  Note that, while we include systematic uncertainties at each redshift, we make the conservative choice to ignore correlations between them at different redshifts.  Thus, the true uncertainties may still be somewhat larger.

\begin{table}
\begin{center}
\caption{Model Fits to Measurement of $\Gamma$ from $z=2.4$--4.75}
\begin{tabular}{ccc|c|c}
\hline
$ \mabs/\msun$ & $\memit/\msun$ & $\fagn$ & $\fesc$ ($\%$) & reduced-$\chi^2$ \\
\hline
$10^{8}$ & $10^{8}$ & 0.25 & 2.8 & 1.5 \\ 
$10^{9}$ & $10^{9}$ & 0.25 & 1.6 & 1.5 \\ 
$10^{10}$ & $10^{10}$ & 0.25 & 1.0 & 1.9 \\
$\mf$ & $\mf$ & 0.25 & 1.6 & 3.5 \\ 
\hline
$10^{8}$ & $10^{8}$ & 1 & 2.1 & 7.6 \\
$10^{9}$ & $10^{9}$ & 1 & 1.0 & 8.0 \\
$10^{10}$ & $10^{10}$ & 1 & 0.5 & 17.1 \\
$\mf$ & $\mf$ & 1 & 1.0 & 20.4 \\ 
\hline
$10^{8}$ & $10^{9}$ & 0.25 & 3.0 & 2.1 \\ 
$10^{8}$ & $10^{10}$ & 0.25 & 3.4 & 4.2 \\
$10^{8}$ & $10^{11}$ & 0.25 & 4.9 & 15.8 \\ 
\hline
\end{tabular}\label{tab:results}
\begin{list}{}{}
{A comparison among the different models we consider in this work of $\chi^2$ fits to the \citet{BB13} measurements of $\Gamma$ only.  The minimum halo mass of absorbers and emitters, $\mabs$ and $\memit$, respectively, and the included fraction, $\fagn$, of the \citet{HM12} AGN emission are held fixed, while the ionizing escape fraction from star formation, $\fesc$, is adjusted to minimize values of the reduced-$\chi^2$.  Values of reduced-$\chi^2$ near unity are considered good fits to the data.}
\end{list}
\end{center}
\end{table}

To compare our model for the gas in halos to the flat evolution in $\Gamma$ represented by the \citet{BB13} observations, we fit a mass- and redshift-independent value of $\fesc$ using the covariance matrix given in Appendix~\ref{sec:app:fit} over the redshift range from $z=2.4$--4.75, the range over which we expect absorbing gas to be confined to halos.  We additionally set $\lion=2.7\times10^{27}\,\erg\,\s^{-1}\,\Hz^{-1}\,\msun^{-1}\,\yr$, $\fagn=0.25$, and $\mabs=\memit$ in this section and defer exploring the effects of increasing the AGN contribution or decoupling $\mabs$ and $\memit$ to \S\ref{sec:assump:agn} and \S\ref{sec:assump:equiv}.

We present the reduced-$\chi^2$ values for each choice of $\mabs=\memit$ in Table \ref{tab:results}.  Values close to unity indicate the best fits.  Moreover, in the left-hand column of Figure~\ref{fig:abs}, we plot the resulting mean free paths, ionizing emissivities, and background ionization rates in the top, middle, and bottom panels, respectively.  All models fit the \citet{BB13} data extremely well, demonstrating that the coupling of absorbers and sources inside halos can account for the flat evolution in $\Gamma$ without invoking evolution in $\fesc$.

We can additionally compare our results for the mean free path to those inferred from recent observations of quasar absorption lines.  Note that the measured mean free path in the literature is not identical to that given in equation~\ref{eq:mfp}.  Instead, the appropriate value for comparison is \citep[see][for a discussion]{BB13}
\begin{equation}\label{eq:mfpobs}
\lambda_{\rm mfp}^{\rm obs}(z_{2}) =  \int_{z_{1}}^{z_{2}}\! \dd z\,\frac{\dd l}{\dd z},
\end{equation}
where $z_{2}$ is the measurement redshift and
\begin{equation}
\int_{z_1}^{z_2}\! \dd z'  \, \frac{\dd \tau_{\rm eff}[\nu_{912}\,(1+z')/(1+z_2),z']}{\dd z'} = 1.
\end{equation}
We compare our results to data from \citet{OMeara13} at $z=2.44$, \citet{Fumagalli13} at $3.00$, \citet{Prochaska09} at $z=3.73$--$4.22$, and \citet{Worseck14} at $z=4.56$--$5.16$ as compiled by \citet{Worseck14}.\footnote{We ignore the slight difference in cosmological parameters between this paper and observational works in the literature, which typically take $\Omega_{\rm m}=0.3$ rather than $0.28$.}  Additionally, we over-plot independent derivations from \citet{Bolton05} and \citet{BH07}.

We find that models with constant minimum masses produce somewhat shallower evolution in the mean free path than observed.  Lower values of $\mabs=\memit$ under-predict $\mfp$ at lower redshifts, while higher values over-predict $\mfp$ at higher redshifts.  On the other hand, the model in which the minimum mass for gas and star formation is set by the filtering mass maintains the flat evolution in $\Gamma$ (with a best-fit value of $\fesc=0.018$ and only a slightly higher reduced-$\chi^2$) while simultaneously producing much closer agreement between the predicted $\mfp(z)$ and observations.  Of course, the required evolution in the minimum mass is theoretically expected and physically motivated by a combination of the ionizing background, the Jeans instability, and heating and cooling \citep[e.g.,][]{Gnedin00, BL01, Hoeft06, Okamoto08, Naoz09, NM14}.  Qualitatively, as the universe grows less dense at lower redshifts, larger halo masses are required to retain gas.  Thus, we can interpret the steep evolution of $\mfp(z)$ as owing partially to these effects.  However, as we noted in \S\ref{sec:sam:abs:det}, the filtering mass ignores the detailed halo formation histories that \citet{NM14} show are important for computing the minimum mass.  We therefore ascribe most of the remaining deviation between our model using $\mf$ and the $\mfp$ measurements at $z\sim4$--5 to the approximate nature of equation~\ref{eq:mf} but leave a more detailed fitting of all available data to future work.

\subsection{The Drop-off in $\Gamma$ at $z\gtrsim 5$}\label{sec:results:dropoff}

Beyond $z\sim5$, gas well outside the splashback radius begins to dominate the absorption.  Figure~\ref{fig:pdf} shows that, by $z=6$, this gas is more appropriately modeled by the \citet{BB09} simulation of the IGM than by average halo profiles.  To gauge the impact of diffuse, intergalactic gas more quantitatively, we compute the evolution in $\Gamma$ using the composite PDF suggested by Figure~\ref{fig:pdf} and proposed in \S\ref{sec:sam:abs}.  That is, at each overdensity, we take the maximum value our PDF using the halo density profile (truncated at $\rsp$) and the simulated \citet{BB09} PDF outside halos.

To calculate the resulting evolution in $\Gamma$, we first employ the simple MHR {\it{ansantz}} for deriving the mean free path from the density PDF:
\begin{equation}\label{eq:pdf-mfp}
\mfp=l_{0}\,\left[\int_{\Di}^{\infty}\!\! {\dd \Delta}\,\frac{\dd P_{\rm V}(\Delta,z)}{\dd \Delta}\right]^{-2/3}, 
\end{equation}
where $l_0(z)=a_0/H(z)$, and we set $a_{0}=95\,\km\,\s^{-1}$ (somewhat higher than the value of $60\,\km\,\s^{-1}$ chosen by MHR for a different set of cosmological parameters).  To additionally account for any contribution to absorption from gas below $10^{17}\,\cm^{-2}$ \citep[e.g.,][]{HM12}, we use $\NLLS=5\times10^{16}\,\cm^{-2}$ in equation~\ref{eq:Di} to compute $\Di$.  

We can then compute $\Gamma$ using \citep[e.g.][]{SB03, CAFG08a}:
\begin{equation}\label{eq:Gamma1}
\Gamma(z)=\frac{\sigma_{\HI}\,\eion\,\mfp}{h_{\rm p}\,(\alpha_{912}+3)}\,(1+z)^3,
\end{equation}
with $\alpha_{912}=2$ and where the factor of $(1+z)^3$ converts the comoving emissivity $\eion$ into physical units.  Equation~\ref{eq:Gamma1} is an approximate version of equation~\ref{eq:Gamma} and ignores cosmological radiative transfer effects that can produce slight over-estimates in $\Gamma$ at $z\lesssim3.5$.  Nevertheless, it is useful for our purposes here, in the absence of a column density distribution function, and produces reliable results through the transition redshift from the halo profile part to the \citet{BB09} part of our composite PDF.  

Deriving $\eion$ from our emitter model in \S\ref{sec:sam:emit} with $\fesc=0.03$ and inserting equation~\ref{eq:pdf-mfp} into equation~\ref{eq:Gamma1}, we iterate until convergence.  Note that, while values of $a_{0}$, $\NLLS$, and $\fesc$ are all reasonable choices selected to produce good agreement between these results and the observations of both $\Gamma$ and $\mfp$, we did not perform a more careful fit because of the approximate nature of equations~\ref{eq:pdf-mfp} and~\ref{eq:Gamma1}.

The right-hand column of Figure~\ref{fig:abs} compares our fiducial halo calculation to results derived from the composite PDF.  Because the diffuse IGM is not correlated with the emissivity of the galaxies inside the halos, the background ionization rate begins to evolve rapidly with $\eion$, decreasing steeply toward higher redshifts, and we recover the scenario investigated by \citet{McQuinn11}.  Furthermore, this simple calculation predicts a break in the evolution of the mean free path from its power-law behavior at $z\lesssim5$ \citep{Worseck14} toward a steeper decline at $z\gtrsim5$, despite no sharp change in the emissivity. Note that the emissivities derived for the composite PDF are comparable (to within a factor of 2) to that of the halo-based model. Given the imprecision in the MHR mean free path ansatz (equation \ref{eq:pdf-mfp}), and other model assumptions such as the cut-off column density, this is acceptable agreement.

The figure also compares the composite PDF results to measurements of the background ionization rate by \citet{WB11} and \citet{Calverley11} who both find values at $z=6$ nearly an order of magnitude lower than do \citet{BB13} at $z=4.75$.  The agreement between these data and our simple model is excellent.\footnote{On the other hand, \citet{Becker14} suggest that the ionization rate at $z\gtrsim5$ may also be significantly patchier than at lower redshifts, potentially increasing the uncertainties on these measurements.}  Thus, contrary to previous claims, the sharp decline in the ionizing background need not signal the end of reionization but only a change in the coupling between sources and absorbers.

\section{Sensitivity to Model Assumptions}\label{sec:assump}

\subsection{The AGN Contribution}\label{sec:assump:agn}

\begin{figure}
\begin{center}
\includegraphics[width=\columnwidth,trim=48 60 305 10,clip]{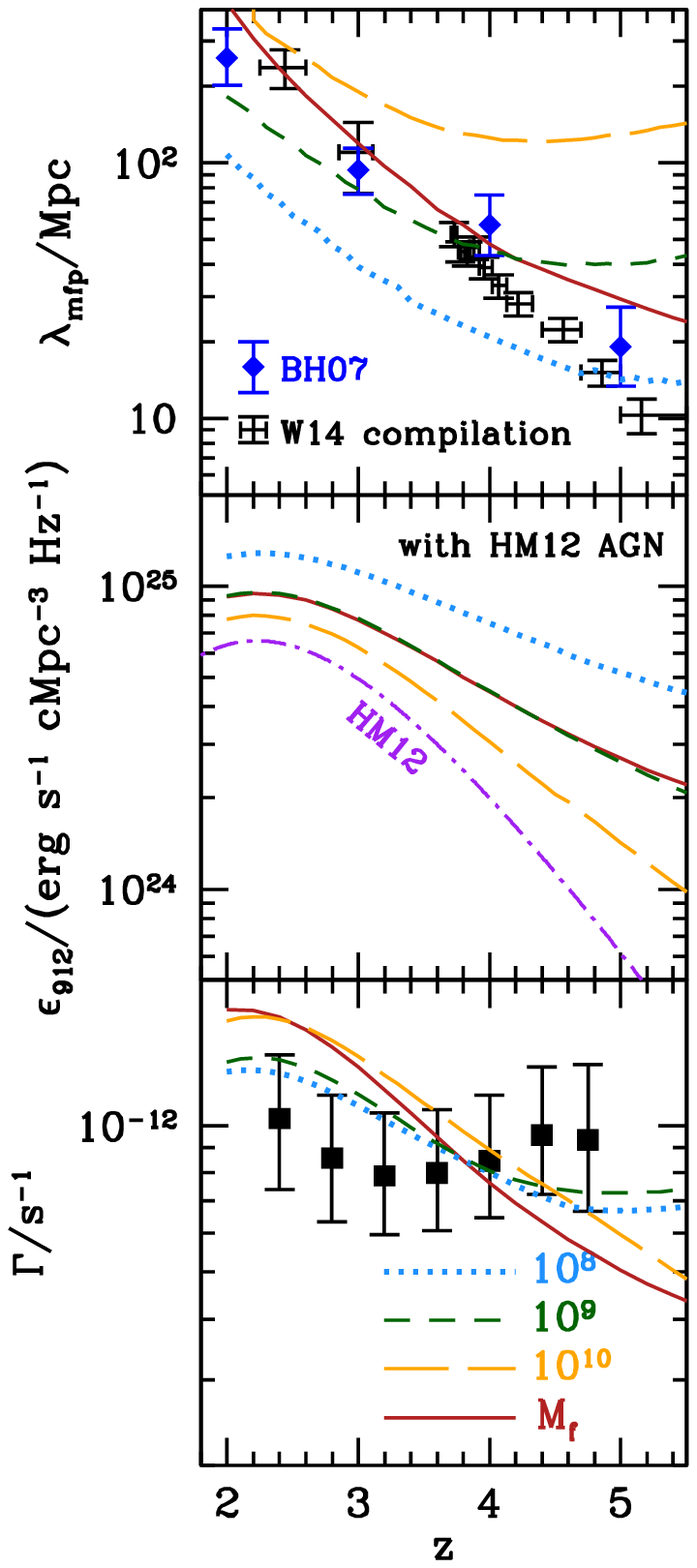}
\caption{\label{fig:agn}  
Same as the left-hand column of Figure~\ref{fig:abs} except that we now set $\fagn=1$ in our models to add the contribution to the emissivity from AGN estimated in \citet[][HM12]{HM12}, which is additionally denoted by the labeled dot-dashed (purple) curve.
}
\end{center}
\end{figure}

In \S\ref{sec:results:flatness}, we show that the coupling between absorbing gas in halos and galactic sources of ionizing radiation within halos can explain the flat evolution in $\Gamma$ observed by \citet{BB13}.  However, absorbers and sources are no longer coupled if the ionizing emissivity is dominated by AGN, which are hosted only by rare, massive halos.  Though we argue in \S\ref{sec:sam:emit:agn} that the contribution from AGN is likely negligible, here we quantitatively examine the effect of an important AGN component by adopting the emissivity model of \citet[][i.e., by setting $\fagn=1$]{HM12}.

We perform the same fits as in \S\ref{sec:results:flatness} for the same choices of $\mabs=\memit$, tabulate the resulting values of $\fesc$ and reduced-$\chi^2$ in Table~\ref{tab:results}, and plot the resulting mean free paths, ionizing emissivities, and background ionization rates in Figure~\ref{fig:agn}.  As expected, the increased steepness of $\eion$, with no corresponding change in the absorber population, ultimately produces a more rapidly evolving $\Gamma$ that, given the high reduced-$\chi^2$ values, is inconsistent with the observations for a non-evolving $\fesc$.

\subsection{The $\mabs$-$\memit$ Equivalence}\label{sec:assump:equiv}

\begin{figure}
\begin{center}
\includegraphics[width=\columnwidth,trim=5 20 10 0,clip]{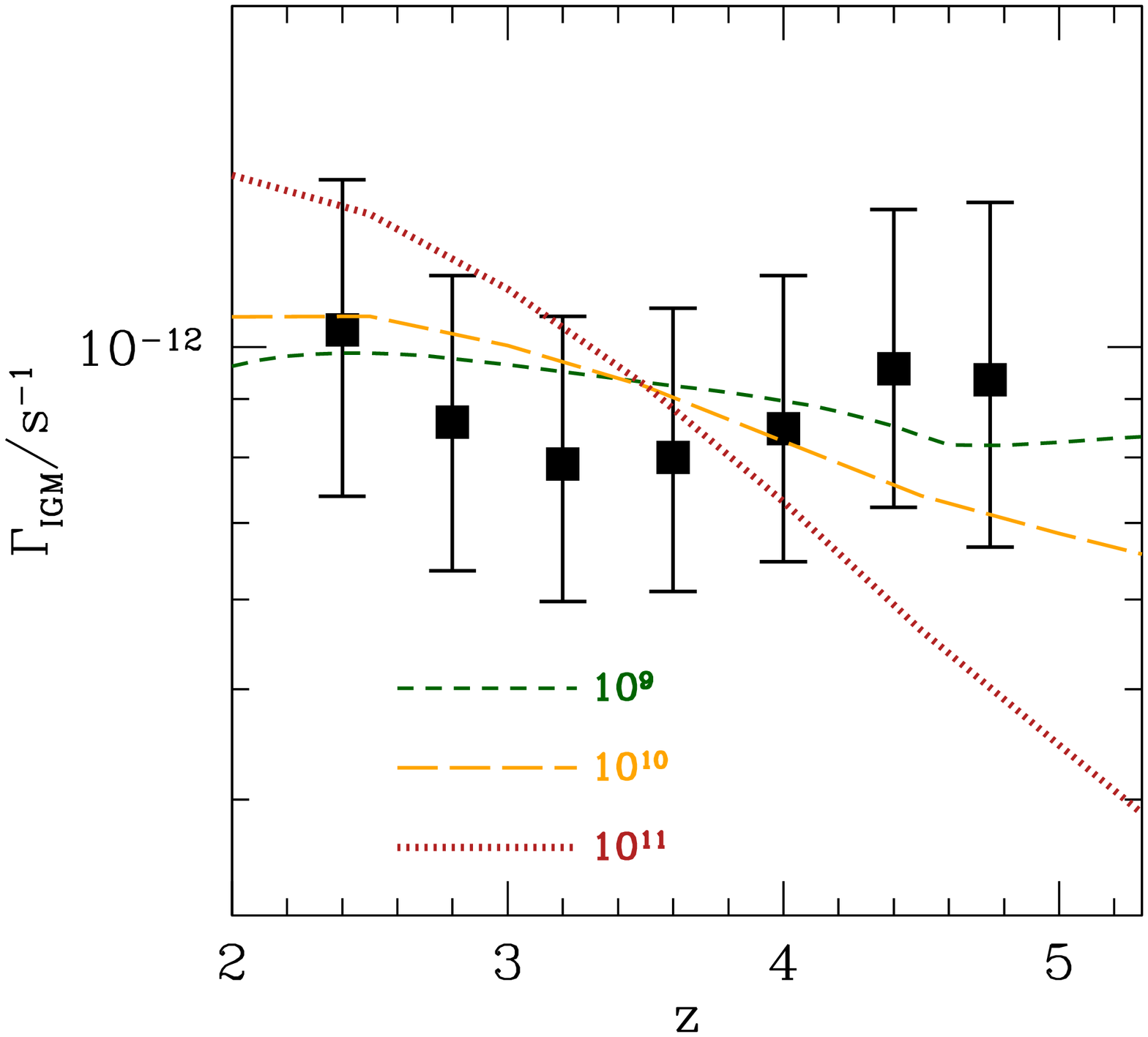}
\caption{\label{fig:equiv} 
The evolution of the ionization rate for $\mabs<\memit$.  $\log\mabs/\msun=8$ in all cases.  We show results for $\log\memit/\msun=9$ (short-dashed, green), $10$ (long-dashed, orange), and $11$ (dotted, red).  The points are the measurements from \citet{BB13}, and the $\chi^2$ values when compared to this data are listed in Table~\ref{tab:results}.
}
\end{center}
\end{figure}

The flat evolution of the background ionization rate which we derive from $z\sim 2$--$5$ rests on the association between sources and absorbers through the formation of cosmic structure, and in our model, we have so far set $\mabs=\memit$ to reflect this.  However, these two minimum masses need not be precisely equal to maintain the connection between sources and absorbers and produce a constant ionizing background.  Physically, this scenario may result from the suppression of star formation due to low metallicity and corresponding low molecular fractions in galaxies that are otherwise be able to retain their gas \citep[e.g.,][]{KD12}.  As a test of this effect on $\Gamma$, we fix $\log\mabs/\msun=8$ and consider progressively larger values of $\memit$, each of which is also held constant in time.  The results are summarized in Table~\ref{tab:results} and plotted in Figure~\ref{fig:equiv}.  The fit to the \citet{BB13} measurements only becomes intolerable if $\log\memit/\msun \gtrsim 10$.  This is because both absorption and emission are completely dominated by halos near the minimum mass only if the minimum mass is above the knee of the mass function, i.e., the non-linear mass $M_{\rm NL}$.\footnote{The non-linear mass is the smoothing mass for which the variance of density fluctuations in the universe is approximately unity \cite[see, e.g.,][]{BL01}.}  If, instead, both minimum masses are below this non-linear mass, as is the case when we set $\mabs=\memit=\mf$, then there will be a contribution to both absorption and emission from halos near $M_{\rm NL}$ that will keep the two processes linked regardless of the specific values of $\mabs$ and $\memit$.  Thus, since $M_{\rm NL}(z=2)\approx 5\times10^{11}\,\msun$ and $M_{\rm NL}(z=5)\approx 10^{9}\,\msun$, our results are insensitive to the precise equivalence between the minimum masses for sources and absorbers.

\subsection{The Shape of the Density Profile and the Self-Shielding Prescription}\label{sec:assump:profile}

In our model, the distribution of neutral hydrogen around galaxies as a function of host halo mass and redshift is a combination of our assumed NFW density profile for gas in halos as well as our ionization prescription with its implementation of gas self-shielding.  However, there can be strong fluctuations in the shape of the dark matter profile about the analytic average in equation~\ref{eq:nfw} \citep[e.g.,][]{DK14}, and the distribution of gas relative to the dark matter is complicated by accretion mode, cooling, and feedback \citep[e.g.,][]{CAFG14}.  Moreover, equations~\ref{eq:Gloc} and~\ref{eq:nss}, which encapsulate our self-shielding prescription, represent only approximate fits of a complicated radiative-transfer process.  

Yet, the flatness in $\Gamma(z)$ is insensitive to all of these details.  This is because, in our model, the dominant contributions to the evolution of the mean free path are the expansion of the universe, the evolution of the filtering mass, and the addition of new halos through the growth of structure rather than a change in the absorber cross-section.  Heuristically, if absorbers are of order the Jeans length \citep{Schaye01}, then the lack of evolution in the physical density for self-shielding (equation~\ref{eq:nss}) implies that the size of absorbers is also roughly redshift independent.  To see this in detail, consider the evolution of $\rLLS/\rvir$ in Figure~\ref{fig:rLLS}.  $\rLLS/\rvir$ increases by just over a factor of $2$ from $z=2$--$4.5$, while the virial radius, which is approximately proportional to $\mhalo^{1/3}\,(1+z)^{-1}$ \citep{BL01}, decreases by nearly the same factor.  As a result, at a given halo mass, $\rLLS$ evolves very little, and the evolution in the mean free path is linked to the growth of structure and the changing abundance of sources.  Therefore, the details that effectively control $\rLLS$ do not strongly influence our result.  

One could imagine implementing a very different model for the profile of density or ionization state around halos that does produce significant evolution in $\rLLS$.  However, such a model would still have to account for the inevitable cosmological effects on the absorber population.  Since our current framework is already consistent with observations of $\mfp(z)$, a model that additionally includes a rapidly evolving $\rLLS$ would likely break this agreement.

Finally, we have assumed that only the external background is important for photoionization balance and ignored the local radiation field.  While recent simulations examining the influence of local photoionization found a potentially significant contribution for damped Lyman-$\alpha$ systems, they determined that the effect on LLSs is negligible \citep{Rahmati13b} in agreement with previous analytic estimates \citep{MiraldaEscude05, Schaye06}.

\section{Conclusions}\label{sec:conc}

We have shown that a framework in which both neutral absorbing gas and the sources of ionizing radiation are associated with the same population of dark matter halos and linked to the growth of cosmic structure generically produces a flat evolution in the background ionization rate from $z \sim 2$--$5$ as measured from quasar absorption lines \citep{BB13}.  Analytically, $\Gamma$ is approximately proportional to $(\epsilon/\na)^{3}$ rather than to $\epsilon^{3}$ so that, for fixed $\fesc$, an increase in the emissivity (in the comoving frame) is compensated for by an increase in the abundance of absorbers.  Indeed, the result of a non-evolving $\Gamma$ is largely independent of our detailed assumptions about the minimum halo mass that supports absorbers and sources, the shape of the density profile around halos, and the self-shielding of neutral gas, though it does require that galaxies dominate the ionizing emissivity.  Moreover, adopting a minimum halo mass for absorbers and emitters which evolves in a way consistent with theoretical expectations, our model also roughly reproduces measurements of $\mfp(z)$.  However, the relationship between sources and absorbers breaks down at still higher redshifts when the mean density of the universe is large enough that gas outside halos must contribute significantly to the absorption of ionizing radiation.  At this point, the background ionization rate becomes extremely sensitive to the source emissivity, as suggested by \citet{McQuinn11}, and $\Gamma$ drops precipitously at $z\sim5$--$6$, consistent with observations \citep{WB11, Calverley11}.  Thus, our model presents a generic solution to the puzzling flatness and sudden evolution of the ionizing background that does not require fine-tuning of the ionizing escape fraction.

The basic association between sources and absorbers in our model can be tested observationally by cross-correlating LLSs in quasar spectra and catalogs of faint galaxies or between LLSs and damped Lyman-$\alpha$ absorbers \citep[e.g.,][]{Font-Ribera12}.  However, in detail, such tests will depend on assumptions about the gas profile of dark matter halos, preferably implemented in numerical simulations.  We leave a more in depth study of different possible halo profiles and configurations and the resulting observable signatures to future work.

In addition to explaining the flat evolution of the background ionization rate, our model reveals new insights into both the production and absorption of ionizing photons.  First, to fit our model to observations of $\Gamma(z)$, we generically require $\fesc$ of order a couple percent, roughly consistent with direct measurements \citep{Nestor13, Jones13b, Mostardi13, Mostardi15}. 
Moreover, the connection between LLSs and halos in our model implies that metal lines observed in quasar absorption spectra and associated with $\HI$ column densities $\gtrsim 10^{17}\,\cm^{-2}$ at $z\lesssim5$ probe the circumgalactic medium around galaxies rather than true intergalactic gas.  However, our results suggest that these same metal lines at even higher redshifts are more likely to be true tracers of the IGM \citep[see, e.g.,][]{Simcoe12, Finlator13}.
Finally, our model provides cosmological context for the evolution of the mean free path, which we attribute to inevitable cosmological processes---a combination of (a) the expansion of the universe, (b) the evolution of the filtering halo mass below which accretion is suppressed, and (c) the changing abundance of halos---without requiring the changes in absorber size, mass, or ionization fraction suggested by \citet{Worseck14}.

An association between sources and absorbers of ionizing radiation is quickly becoming canonical.  If true, this idea will link future observations of the background ionization rate and quasar absorption lines, not only to the star formation in galaxies, but to their gas and halo structure as well.

\section{Acknowledgements}
We thank George Becker, James Bolton, Piero Madau, and the referee for helpful insights.  JAM and SPO acknowledge NASA grant NNX12AG73G for support.

\bibliography{ms_rev2}

\begin{thebibliography}{78}
\expandafter\ifx\csname natexlab\endcsname\relax\def\natexlab#1{#1}\fi

\bibitem[{{Barkana} \& {Loeb}(2001)}]{BL01}
{Barkana} R., {Loeb} A., 2001, \physrep, 349, 125

\bibitem[{{Becker} \& {Bolton}(2013)}]{BB13}
{Becker} G.~D., {Bolton} J.~S., 2013, \mnras

\bibitem[{{Becker} {et~al}\mbox{.}(2014){Becker}, {Bolton}, {Madau}, {Pettini},
  {Ryan-Weber}, \& {Venemans}}]{Becker14}
{Becker} G.~D., {Bolton} J.~S., {Madau} P., {Pettini} M., {Ryan-Weber} E.~V.,
  {Venemans} B.~P., 2014, astro-ph/1407.4850

\bibitem[{{Becker} {et~al}\mbox{.}(2013){Becker}, {Hewett}, {Worseck}, \&
  {Prochaska}}]{Becker13}
{Becker} G.~D., {Hewett} P.~C., {Worseck} G., {Prochaska} J.~X., 2013, \mnras,
  430, 2067

\bibitem[{{Becker}, {Rauch} \& {Sargent}(2007){Becker}, {Rauch}, \&
  {Sargent}}]{Becker07}
{Becker} G.~D., {Rauch} M., {Sargent} W.~L.~W., 2007, \apj, 662, 72

\bibitem[{{Becker} \& {Kravtsov}(2011)}]{BK11}
{Becker} M.~R., {Kravtsov} A.~V., 2011, \apj, 740, 25

\bibitem[{{Bolton} \& {Becker}(2009)}]{BB09}
{Bolton} J.~S., {Becker} G.~D., 2009, \mnras, 398, L26

\bibitem[{{Bolton} \& {Haehnelt}(2007)}]{BH07}
{Bolton} J.~S., {Haehnelt} M.~G., 2007, \mnras, 382, 325

\bibitem[{{Bolton} {et~al}\mbox{.}(2005){Bolton}, {Haehnelt}, {Viel}, \&
  {Springel}}]{Bolton05}
{Bolton} J.~S., {Haehnelt} M.~G., {Viel} M., {Springel} V., 2005, \mnras, 357,
  1178

\bibitem[{{Bouwens} {et~al}\mbox{.}(2007){Bouwens}, {Illingworth}, {Franx}, \&
  {Ford}}]{Bouwens07}
{Bouwens} R.~J., {Illingworth} G.~D., {Franx} M., {Ford} H., 2007, \apj, 670,
  928

\bibitem[{{Bouwens} {et~al}\mbox{.}(2012){Bouwens}, {Illingworth}, {Oesch},
  {Trenti}, {Labb{\'e}}, {Franx}, {Stiavelli}, {Carollo}, {van Dokkum}, \&
  {Magee}}]{Bouwens12}
{Bouwens} R.~J. {et~al.}, 2012, \apjl, 752, L5

\bibitem[{{Calverley} {et~al}\mbox{.}(2011){Calverley}, {Becker}, {Haehnelt},
  \& {Bolton}}]{Calverley11}
{Calverley} A.~P., {Becker} G.~D., {Haehnelt} M.~G., {Bolton} J.~S., 2011,
  \mnras, 412, 2543

\bibitem[{{Cowie}, {Barger} \& {Trouille}(2009){Cowie}, {Barger}, \&
  {Trouille}}]{Cowie09}
{Cowie} L.~L., {Barger} A.~J., {Trouille} L., 2009, \apj, 692, 1476

\bibitem[{{Cuesta} {et~al}\mbox{.}(2008){Cuesta}, {Prada}, {Klypin}, \&
  {Moles}}]{Cuesta08}
{Cuesta} A.~J., {Prada} F., {Klypin} A., {Moles} M., 2008, \mnras, 389, 385

\bibitem[{{Dav{\'e}}, {Finlator} \& {Oppenheimer}(2012){Dav{\'e}}, {Finlator},
  \& {Oppenheimer}}]{Dave12}
{Dav{\'e}} R., {Finlator} K., {Oppenheimer} B.~D., 2012, \mnras, 421, 98

\bibitem[{{Diemer} \& {Kravtsov}(2014)}]{DK14}
{Diemer} B., {Kravtsov} A.~V., 2014, \apj, 789, 1

\bibitem[{{Dutton} \& {Macci{\`o}}(2014)}]{DM14}
{Dutton} A.~A., {Macci{\`o}} A.~V., 2014, \mnras, 441, 3359

\bibitem[{{Fan} {et~al}\mbox{.}(2006){Fan}, {Strauss}, {Becker}, {White},
  {Gunn}, {Knapp}, {Richards}, {Schneider}, {Brinkmann}, \&
  {Fukugita}}]{Fan06a}
{Fan} X. {et~al.}, 2006, \aj, 132, 117

\bibitem[{{Faucher-Giguere} {et~al}\mbox{.}(2014){Faucher-Giguere}, {Hopkins},
  {Keres}, {Muratov}, {Quataert}, \& {Murray}}]{CAFG14}
{Faucher-Giguere} C.-A., {Hopkins} P.~F., {Keres} D., {Muratov} A.~L.,
  {Quataert} E., {Murray} N., 2014, astro-ph/1409.1919

\bibitem[{{Faucher-Gigu{\`e}re}
  {et~al}\mbox{.}(2008{\natexlab{a}}){Faucher-Gigu{\`e}re}, {Lidz},
  {Hernquist}, \& {Zaldarriaga}}]{CAFG08a}
{Faucher-Gigu{\`e}re} C.-A., {Lidz} A., {Hernquist} L., {Zaldarriaga} M.,
  2008{\natexlab{a}}, \apjl, 682, L9

\bibitem[{{Faucher-Gigu{\`e}re}
  {et~al}\mbox{.}(2008{\natexlab{b}}){Faucher-Gigu{\`e}re}, {Lidz},
  {Hernquist}, \& {Zaldarriaga}}]{CAFG08b}
{Faucher-Gigu{\`e}re} C.-A., {Lidz} A., {Hernquist} L., {Zaldarriaga} M.,
  2008{\natexlab{b}}, \apj, 688, 85

\bibitem[{{Ferrara} \& {Loeb}(2013)}]{FL13}
{Ferrara} A., {Loeb} A., 2013, \mnras, 431, 2826

\bibitem[{{Finlator} {et~al}\mbox{.}(2013){Finlator}, {Mu{\~n}oz},
  {Oppenheimer}, {Oh}, {{\"O}zel}, \& {Dav{\'e}}}]{Finlator13}
{Finlator} K., {Mu{\~n}oz} J.~A., {Oppenheimer} B.~D., {Oh} S.~P., {{\"O}zel}
  F., {Dav{\'e}} R., 2013, \mnras

\bibitem[{{Font-Ribera} {et~al}\mbox{.}(2012){Font-Ribera},
  {Miralda-Escud{\'e}}, {Arnau}, {Carithers}, {Lee}, {Noterdaeme}, {P{\^a}ris},
  {Petitjean}, {Rich}, {Rollinde}, {Ross}, {Schneider}, {White}, \&
  {York}}]{Font-Ribera12}
{Font-Ribera} A. {et~al.}, 2012, \jcap, 11, 59

\bibitem[{{Fumagalli} {et~al}\mbox{.}(2013){Fumagalli}, {O'Meara}, {Prochaska},
  \& {Worseck}}]{Fumagalli13}
{Fumagalli} M., {O'Meara} J.~M., {Prochaska} J.~X., {Worseck} G., 2013, \apj,
  775, 78

\bibitem[{{Furlanetto} \& {Oh}(2005)}]{FO05}
{Furlanetto} S.~R., {Oh} S.~P., 2005, \mnras, 363, 1031

\bibitem[{{Gnedin}(2000)}]{Gnedin00}
{Gnedin} N.~Y., 2000, \apj, 542, 535

\bibitem[{{Haardt} \& {Madau}(2012)}]{HM12}
{Haardt} F., {Madau} P., 2012, \apj, 746, 125

\bibitem[{{Hayashi} \& {White}(2008)}]{HW08}
{Hayashi} E., {White} S.~D.~M., 2008, \mnras, 388, 2

\bibitem[{{Hoeft} {et~al}\mbox{.}(2006){Hoeft}, {Yepes}, {Gottl{\"o}ber}, \&
  {Springel}}]{Hoeft06}
{Hoeft} M., {Yepes} G., {Gottl{\"o}ber} S., {Springel} V., 2006, \mnras, 371,
  401

\bibitem[{{Hopkins} \& {Beacom}(2006)}]{HB06}
{Hopkins} A.~M., {Beacom} J.~F., 2006, \apj, 651, 142

\bibitem[{{Hopkins}, {Richards} \& {Hernquist}(2007){Hopkins}, {Richards}, \&
  {Hernquist}}]{Hopkins07}
{Hopkins} P.~F., {Richards} G.~T., {Hernquist} L., 2007, \apj, 654, 731

\bibitem[{{Jones} {et~al}\mbox{.}(2013){Jones}, {Ellis}, {Schenker}, \&
  {Stark}}]{Jones13b}
{Jones} T.~A., {Ellis} R.~S., {Schenker} M.~A., {Stark} D.~P., 2013, \apj, 779,
  52

\bibitem[{{Krumholz} \& {Dekel}(2012)}]{KD12}
{Krumholz} M.~R., {Dekel} A., 2012, \apj, 753, 16

\bibitem[{{Kuhlen} \& {Faucher-Gigu{\`e}re}(2012)}]{KFG12}
{Kuhlen} M., {Faucher-Gigu{\`e}re} C.-A., 2012, \mnras, 423, 862

\bibitem[{{McBride}, {Fakhouri} \& {Ma}(2009){McBride}, {Fakhouri}, \&
  {Ma}}]{McBride09}
{McBride} J., {Fakhouri} O., {Ma} C.-P., 2009, \mnras, 398, 1858

\bibitem[{{McQuinn}, {Oh} \& {Faucher-Gigu{\`e}re}(2011){McQuinn}, {Oh}, \&
  {Faucher-Gigu{\`e}re}}]{McQuinn11}
{McQuinn} M., {Oh} S.~P., {Faucher-Gigu{\`e}re} C.-A., 2011, \apj, 743, 82

\bibitem[{{Miralda-Escud{\'e}}(2005)}]{MiraldaEscude05}
{Miralda-Escud{\'e}} J., 2005, \apjl, 620, L91

\bibitem[{{Miralda-Escud{\'e}} {et~al}\mbox{.}(1996){Miralda-Escud{\'e}},
  {Cen}, {Ostriker}, \& {Rauch}}]{MiraldaEscude96}
{Miralda-Escud{\'e}} J., {Cen} R., {Ostriker} J.~P., {Rauch} M., 1996, \apj,
  471, 582

\bibitem[{{Miralda-Escud{\'e}}, {Haehnelt} \& {Rees}(2000){Miralda-Escud{\'e}},
  {Haehnelt}, \& {Rees}}]{MiraldaEscude00}
{Miralda-Escud{\'e}} J., {Haehnelt} M., {Rees} M.~J., 2000, \apj, 530, 1

\bibitem[{{More}, {Diemer} \& {Kravtsov}(2015){More}, {Diemer}, \&
  {Kravtsov}}]{More15}
{More} S., {Diemer} B., {Kravtsov} A., 2015, astro-ph:1504.05591

\bibitem[{{Mostardi} {et~al}\mbox{.}(2013){Mostardi}, {Shapley}, {Nestor},
  {Steidel}, {Reddy}, \& {Trainor}}]{Mostardi13}
{Mostardi} R.~E., {Shapley} A.~E., {Nestor} D.~B., {Steidel} C.~C., {Reddy}
  N.~A., {Trainor} R.~F., 2013, \apj, 779, 65

\bibitem[{{Mostardi} {et~al}\mbox{.}(2015){Mostardi}, {Shapley}, {Steidel},
  {Trainor}, {Reddy}, \& {Siana}}]{Mostardi15}
{Mostardi} R.~E., {Shapley} A.~E., {Steidel} C.~C., {Trainor} R.~F., {Reddy}
  N.~A., {Siana} B., 2015, astro-ph/1506.08201

\bibitem[{{Mu{\~n}oz}(2012)}]{Munoz12}
{Mu{\~n}oz} J.~A., 2012, \jcap, 4, 15

\bibitem[{{Mu{\~n}oz} \& {Loeb}(2011)}]{ML11}
{Mu{\~n}oz} J.~A., {Loeb} A., 2011, \apj, 729, 99

\bibitem[{{Murakami} \& {Ikeuchi}(1990)}]{MI90}
{Murakami} I., {Ikeuchi} S., 1990, \pasj, 42, L11

\bibitem[{{Naoz}, {Barkana} \& {Mesinger}(2009){Naoz}, {Barkana}, \&
  {Mesinger}}]{Naoz09}
{Naoz} S., {Barkana} R., {Mesinger} A., 2009, \mnras, 399, 369

\bibitem[{{Navarro}, {Frenk} \& {White}(1997){Navarro}, {Frenk}, \&
  {White}}]{NFW97}
{Navarro} J.~F., {Frenk} C.~S., {White} S.~D.~M., 1997, \apj, 490, 493

\bibitem[{{Nestor} {et~al}\mbox{.}(2013){Nestor}, {Shapley}, {Kornei},
  {Steidel}, \& {Siana}}]{Nestor13}
{Nestor} D.~B., {Shapley} A.~E., {Kornei} K.~A., {Steidel} C.~C., {Siana} B.,
  2013, \apj, 765, 47

\bibitem[{{Noh} \& {McQuinn}(2014)}]{NM14}
{Noh} Y., {McQuinn} M., 2014, \mnras, 444, 503

\bibitem[{{Oguri} \& {Hamana}(2011)}]{OH11}
{Oguri} M., {Hamana} T., 2011, \mnras, 414, 1851

\bibitem[{{Okamoto}, {Gao} \& {Theuns}(2008){Okamoto}, {Gao}, \&
  {Theuns}}]{Okamoto08}
{Okamoto} T., {Gao} L., {Theuns} T., 2008, \mnras, 390, 920

\bibitem[{{O'Meara} {et~al}\mbox{.}(2013){O'Meara}, {Prochaska}, {Worseck},
  {Chen}, \& {Madau}}]{OMeara13}
{O'Meara} J.~M., {Prochaska} J.~X., {Worseck} G., {Chen} H.-W., {Madau} P.,
  2013, \apj, 765, 137

\bibitem[{{Prada} {et~al}\mbox{.}(2006){Prada}, {Klypin}, {Simonneau},
  {Betancort-Rijo}, {Patiri}, {Gottl{\"o}ber}, \& {Sanchez-Conde}}]{Prada06}
{Prada} F., {Klypin} A.~A., {Simonneau} E., {Betancort-Rijo} J., {Patiri} S.,
  {Gottl{\"o}ber} S., {Sanchez-Conde} M.~A., 2006, \apj, 645, 1001

\bibitem[{{Prochaska}, {O'Meara} \& {Worseck}(2010){Prochaska}, {O'Meara}, \&
  {Worseck}}]{Prochaska10}
{Prochaska} J.~X., {O'Meara} J.~M., {Worseck} G., 2010, \apj, 718, 392

\bibitem[{{Prochaska}, {Worseck} \& {O'Meara}(2009){Prochaska}, {Worseck}, \&
  {O'Meara}}]{Prochaska09}
{Prochaska} J.~X., {Worseck} G., {O'Meara} J.~M., 2009, \apjl, 705, L113

\bibitem[{{Rahmati} {et~al}\mbox{.}(2013{\natexlab{a}}){Rahmati}, {Pawlik},
  {Rai{\v c}evi{\'c}}, \& {Schaye}}]{Rahmati13a}
{Rahmati} A., {Pawlik} A.~H., {Rai{\v c}evi{\'c}} M., {Schaye} J.,
  2013{\natexlab{a}}, \mnras, 430, 2427

\bibitem[{{Rahmati} \& {Schaye}(2014)}]{RS14}
{Rahmati} A., {Schaye} J., 2014, \mnras, 438, 529

\bibitem[{{Rahmati} {et~al}\mbox{.}(2013{\natexlab{b}}){Rahmati}, {Schaye},
  {Pawlik}, \& {Rai{\v{c}}evi{\'c}}}]{Rahmati13b}
{Rahmati} A., {Schaye} J., {Pawlik} A.~H., {Rai{\v{c}}evi{\'c}} M.,
  2013{\natexlab{b}}, \mnras, 431, 2261

\bibitem[{{Rauch} {et~al}\mbox{.}(2008){Rauch}, {Haehnelt}, {Bunker}, {Becker},
  {Marleau}, {Graham}, {Cristiani}, {Jarvis}, {Lacey}, {Morris}, {Peroux},
  {R{\"o}ttgering}, \& {Theuns}}]{Rauch08}
{Rauch} M. {et~al.}, 2008, \apj, 681, 856

\bibitem[{{Rauch} \& {Haehnelt}(2011)}]{RH11}
{Rauch} M., {Haehnelt} M.~G., 2011, \mnras, 412, L55

\bibitem[{{Reddy} \& {Steidel}(2009)}]{RS09}
{Reddy} N.~A., {Steidel} C.~C., 2009, \apj, 692, 778

\bibitem[{{Rudie} {et~al}\mbox{.}(2012){Rudie}, {Steidel}, {Trainor}, {Rakic},
  {Bogosavljevi{\'c}}, {Pettini}, {Reddy}, {Shapley}, {Erb}, \&
  {Law}}]{Rudie12}
{Rudie} G.~C. {et~al.}, 2012, \apj, 750, 67

\bibitem[{{Schaye}(2001)}]{Schaye01}
{Schaye} J., 2001, \apj, 559, 507

\bibitem[{{Schaye}(2006)}]{Schaye06}
{Schaye} J., 2006, \apj, 643, 59

\bibitem[{{Schaye} {et~al}\mbox{.}(2003){Schaye}, {Aguirre}, {Kim}, {Theuns},
  {Rauch}, \& {Sargent}}]{Schaye03}
{Schaye} J., {Aguirre} A., {Kim} T.-S., {Theuns} T., {Rauch} M., {Sargent}
  W.~L.~W., 2003, \apj, 596, 768

\bibitem[{{Schirber} \& {Bullock}(2003)}]{SB03}
{Schirber} M., {Bullock} J.~S., 2003, \apj, 584, 110

\bibitem[{{Sheth}, {Mo} \& {Tormen}(2001){Sheth}, {Mo}, \& {Tormen}}]{SMT01}
{Sheth} R.~K., {Mo} H.~J., {Tormen} G., 2001, \mnras, 323, 1

\bibitem[{{Sheth} \& {Tormen}(1999)}]{ST99}
{Sheth} R.~K., {Tormen} G., 1999, \mnras, 308, 119

\bibitem[{{Simcoe} {et~al}\mbox{.}(2012){Simcoe}, {Sullivan}, {Cooksey}, {Kao},
  {Matejek}, \& {Burgasser}}]{Simcoe12}
{Simcoe} R.~A., {Sullivan} P.~W., {Cooksey} K.~L., {Kao} M.~M., {Matejek}
  M.~S., {Burgasser} A.~J., 2012, \nat, 492, 79

\bibitem[{{Sobacchi} \& {Mesinger}(2014)}]{SM14}
{Sobacchi} E., {Mesinger} A., 2014, \mnras, 440, 1662

\bibitem[{{Stark} {et~al}\mbox{.}(2013){Stark}, {Schenker}, {Ellis},
  {Robertson}, {McLure}, \& {Dunlop}}]{Stark13}
{Stark} D.~P., {Schenker} M.~A., {Ellis} R., {Robertson} B., {McLure} R.,
  {Dunlop} J., 2013, \apj, 763, 129

\bibitem[{{Steidel} {et~al}\mbox{.}(2010){Steidel}, {Erb}, {Shapley},
  {Pettini}, {Reddy}, {Bogosavljevi{\'c}}, {Rudie}, \& {Rakic}}]{Steidel10}
{Steidel} C.~C., {Erb} D.~K., {Shapley} A.~E., {Pettini} M., {Reddy} N.,
  {Bogosavljevi{\'c}} M., {Rudie} G.~C., {Rakic} O., 2010, \apj, 717, 289

\bibitem[{{Tavio} {et~al}\mbox{.}(2008){Tavio}, {Cuesta}, {Prada}, {Klypin}, \&
  {Sanchez-Conde}}]{Tavio08}
{Tavio} H., {Cuesta} A.~J., {Prada} F., {Klypin} A.~A., {Sanchez-Conde} M.~A.,
  2008, astro-ph/0807.3027

\bibitem[{{Trenti} {et~al}\mbox{.}(2010){Trenti}, {Stiavelli}, {Bouwens},
  {Oesch}, {Shull}, {Illingworth}, {Bradley}, \& {Carollo}}]{Trenti10}
{Trenti} M., {Stiavelli} M., {Bouwens} R.~J., {Oesch} P., {Shull} J.~M.,
  {Illingworth} G.~D., {Bradley} L.~D., {Carollo} C.~M., 2010, \apjl, 714, L202

\bibitem[{{Worseck} {et~al}\mbox{.}(2014){Worseck}, {Prochaska}, {O'Meara},
  {Becker}, {Ellison}, {Lopez}, {Meiksin}, {M{\'e}nard}, {Murphy}, \&
  {Fumagalli}}]{Worseck14}
{Worseck} G. {et~al.}, 2014, astro-ph/1402.4154

\bibitem[{{Wyithe} \& {Bolton}(2011)}]{WB11}
{Wyithe} J.~S.~B., {Bolton} J.~S., 2011, \mnras, 412, 1926

\bibitem[{{Wyithe} \& {Loeb}(2006)}]{WL06}
{Wyithe} J.~S.~B., {Loeb} A., 2006, \nat, 441, 322

\end{thebibliography}


\begin{appendix}

\setcounter{table}{1}

\begin{table}
\begin{center}
\caption{Covariance Matrix for the Measurements of $\Gamma$ from $z=2.4$--4.75}
\begin{tabular}{c|ccccccc}
\hline
z & 2.4 & 2.8 & 3.2 & 3.6 & 4.0 & 4.4 & 4.75 \\
\hline
2.4 & 1.070 & 0.145 & 0.114 & 0.094 & 0.083 & 0.077 & 0.076 \\
2.8 & & 1.013 & 0.101 & 0.079 & 0.074 & 0.071 & 0.069 \\
3.2 & & & 0.939 & 0.089 & 0.069 & 0.070 & 0.075 \\
3.6 & & & & 0.898 & 0.092 & 0.065 & 0.074 \\
4.0 & & & & & 0.879 & 0.117 & 0.079 \\
4.4 & & & & & & 0.911 & 0.183 \\
4.75 & & & & & & & 1.182 \\
\hline
\end{tabular}\label{tab:cov}
\begin{list}{}{}
{The full covariance matrix for the \citet{BB13} data including statistical, Jeans smoothing, and systematic uncertainties (see text).  Values have been multiplied by a factor of 100 for convenient notation.}
\end{list}
\end{center}
\end{table}

\section{Fitting Observations of the Background Ionization Rate} \label{sec:app:fit}

To compute the symmetric covariance matrix, $\vec{C}$, given in Table~\ref{tab:cov}, we start with the covariance matrix for the statistical uncertainties in $\log\Gamma$ given by Table 2 of \citet{BB13} and add the Jeans smoothing uncertainties and the systematic uncertainties quoted by these authors to all elements and to diagonal elements, respectively.  We then determine our best-fit models by minimizing 
\begin{equation}\label{eq:chi2}
\chi^2=(\vec{x}_{\rm mod}-\vec{x}_{\rm obs})^{\rm T}\,\vec{C}^{-1}\,(\vec{x}_{\rm mod}-\vec{x}_{\rm obs}),
\end{equation}
where $\vec{x}_{\rm obs}$ and $\vec{x}_{\rm mod}$ are vectors containing mean values of $\log\Gamma$, respectively, computed by our model and measured by \citet{BB13} for the same set of redshifts.

\section{Implementation of Outer Density Profile} \label{sec:app:dk}

We adopt the outer halo density profile derived from numerical simulations by \citet{DK14} in which
\begin{equation}
\Delta(r)=f_{\rm trans}\,\Delta_{\rm inner}+\Delta_{\rm outer},\nonumber
\end{equation}
\begin{equation}
f_{\rm trans}=\left[1+\left(\frac{r}{(1.9-\,0.18\nu)\,R_{\rm 200m}}\right)^{4}\right]^{-2},\nonumber
\end{equation}
and
\begin{equation}\label{eq:Douter}
\Delta_{\rm outer}=b_{\rm e}\,\left(\frac{r}{5\,R_{\rm 200m}}\right)^{-s_{\rm e}}+1.
\end{equation}
In equation~\ref{eq:Douter}, $\nu=1.686/\sigma(M, z)$, $\sigma(M, z)$ is the standard deviation of linear density fluctuations on size scales containing mass halo mass $M$ \citep[e.g.,][]{BL01} at redshift $z$, and we set
\begin{equation}
R_{\rm 200m} \approx \rvir \,\left[\frac{\Dvir}{200\,\Omega_{\rm m}\,(1+z)}\right]^{1/3},
\end{equation}
which assumes that the mass enclosed within $R_{\rm 200m}$ is not much different than that within $\rvir$, a reasonable supposition given that both radii are in the outskirts of a halo and that $R_{\rm 200m}$ is relatively insensitive to the ratio of these two masses.  We further adopt the NFW profile from equation~\ref{eq:nfw} for $\Delta_{\rm inner}$ and set $b_{\rm e}=2.0$ and $s_{\rm e}=1.3$ as constant based on Figure~18 of \citet{DK14} and our needs at high redshift.

As noted in \S\ref{sec:sam:abs}, we also truncate the density profile at radii beyond the spashback radius and adopt a fitting function for $\rsp$ derived from simulations by \citet{More15}:
\begin{equation}
\frac{\rsp}{R_{\rm 200m}}=0.81\,\left(1+0.97\,{\rm e}^{-\nu/2.44}\right).
\end{equation}

\end{appendix}

\end{document}